\documentclass[aps,prd,english,superscriptaddress, longbibliography,11pt,notitlepage, nofootinbib]{revtex4}

	\usepackage[colorlinks=true, a4paper=true, pdfstartview=FitV,
linkcolor=blue, citecolor=blue, urlcolor=blue]{hyperref}

\usepackage{amsmath}
\usepackage{amssymb}
\usepackage{graphicx}
\usepackage{hhline}
\usepackage{mwe}
\usepackage{amsbsy}
\usepackage{textcomp}
\usepackage{commath}
\usepackage{subfig}
\usepackage{slashed}
\usepackage{natbib}
\usepackage{xcolor}

\DeclareMathOperator{\sech}{sech}

\usepackage{stackengine}
\stackMath

\makeatletter
\usepackage{babel}
\numberwithin{equation}{section}
\begin{document}
\immediate\write16{<<WARNING: LINEDRAW macros work with emTeX-dvivers
                    and other drivers supporting emTeX \special's
                    (dviscr, dvihplj, dvidot, dvips, dviwin, etc.) >>}

\title{Collision Dynamics of False-Vacuum Oscillons}

\author{J.~G.~F. Campos}
\email{joao.gfcampos@upe.br}
\affiliation{Física de Materiais, Universidade de Pernambuco, Rua
Benfica, 455, Recife - PE - 50720-001, Brazil}

\author{N.~S. Manton}
\email[]{N.S.Manton@damtp.cam.ac.uk}
\affiliation{Department of Applied Mathematics and Theoretical Physics,
University of Cambridge, Wilberforce Road, Cambridge CB3 0WA, U.K.}

\author{A. Mohammadi}
\email[]{azadeh.mohammadi@ufpe.br}
\affiliation{Departamento de Física, Universidade Federal da
Pernambuco, Av. Prof. Moraes Rego, 1235, Recife - PE - 50670-901,
Brazil}

\begin{abstract}
We study the collision dynamics of localized oscillons in two classes of
$(1+1)$-dimensional scalar field theories with metastable false
vacua, a normal class with a positive quartic self-interaction term
and an inverted class with a negative quartic term. We construct
small-amplitude oscillon solutions around the false vacuum using
the Fodor {\emph{et al.}} expansion, and show that the force
between oscillons decays exponentially at large separation, with a
strength modulated by their relative phase. Numerical simulations of
two-oscillon collisions exhibit reflection, crossing, and formation
of excited oscillons. Resonance windows occur, similar to those found
in kink-antikink collisions. In the normal theory, if the oscillons
have sufficient energy, the field can pass over a sphaleron barrier and
evolve into a kink-antikink pair, initiating a phase transition to the
true vacuum. We also simulate the collision of oscillons evolved from a slightly perturbed sphaleron.

\end{abstract}

\maketitle

\section{Introduction}
\label{intro}

Oscillons are among the most ubiquitous classical solutions in
nonlinear field theories \cite{gleiser1994pseudostable,
  copeland1995oscillons}. They describe localized
oscillations of a scalar field around the vacuum, and frequently
occur in realistic field theories. Oscillons are believed to play an
important role in many physical contexts, such as inflationary dynamics
\cite{amin2010inflaton, raphael2010oscillations, lozanov2023enhanced,
  aurrekoetxea2023oscillon}, cosmological phase transitions
\cite{pirvu2024bubble}
 and Bose-Einstein
condensates \cite{charukhchyan2014spatially}. The integrable counterpart
of an oscillon is a breather, a localized solution that oscillates
without emitting radiation.  

Although oscillons arise in non-integrable theories, they are typically
extremely long-lived -- one of their most remarkable and still not fully
understood properties  \cite{graham2006unnatural, salmi2012radiation,
  zhang2020classical, olle2021recipes, cyncynates2021structure,
  van2023oscillon}. While breather solutions are often known
analytically, oscillons can only be constructed perturbatively,
through the Fodor \emph{et al.} asymptotic expansion
\cite{fodor2006oscillons, fodor2008small}, or numerically. It has
recently been shown that oscillons can be mapped onto Q-ball
solutions, providing a novel explanation for their longevity in terms
of Q-ball stability \cite{blaschke2025oscillons}.

The Fodor expansion implies that the oscillon spectrum consists of a
single frequency and its harmonics -- the Fodor
frequencies\footnote{For brevity, we will refer to Fodor
oscillons, rather than Fodor \emph{et al.} oscillons, and sometimes
use just the leading term in their asymptotic expansion.}. The linearized excitations around a small amplitude oscillon solution contain only continuum modes \cite{ fodor2009computation, wang2023excited, evslin2025normal,
evslin2025universal}. Excited oscillons have been shown
to play an important role as intermediate configurations in
kink-antikink \cite{peyrard1983kink} and Q-ball collisions
\cite{alonso2025excited}. Information about their spectrum allows the study
of quantum properties of breathers \cite{dashen1975particle} and oscillons
\cite{evslin2023quantum}. 

Oscillons in (1+1)-dimensions can also be interpreted as bound
kink-antikink pairs, due to the attraction between kinks and
antikinks. When the vacuum surrounding the pair is a local but not global
minimum of the potential (i.e., a false vacuum), a repulsion is also
generated, and at the separation where attraction and repulsion balance,
the system admits a static but unstable solution known as a sphaleron. The
decay of such a sphaleron typically leads in one direction to an
excited oscillon \cite{manton2023simplest}, and in the other
to an expanding kink-antikink pair with true vacuum between them.
The presence of a vibrational mode of the sphaleron
can significantly affect the emerging excited oscillon
\cite{navarro2024impact}. The nature of the false vacuum is also
crucial. If its field fluctuations are massless, leading to
long-range interactions between kinks and antikinks, then oscillon
formation is usually suppressed and a large amount of
radiation is produced instead \cite{campos2021interaction,
  bazeia2023kink}. However, if the field is long-range in a restricted region of the target space, it acquires an effective mass and oscillons form \cite{dorey2024oscillons}. Also, weakly long-range kinks can exhibit resonance windows when scattered \cite{campos2025resonance}.

As oscillons are not precisely understood mathematically, their
dynamics needs to be investigated through numerical simulations.
In ref.~\cite{romanczukiewicz2010oscillon}, the authors simulated
wave-packet (particle-like) collisions and created kink-antikink
pairs via oscillon excitation. Recently, excited oscillons arising
from a Gaussian wave packet have been studied and the creation of
oscillon pairs observed in several multi-bounce windows
\cite{blaschke2024amplitude,simas2025vacuum}. Scattering of
compactly-supported oscillons of the signum-Gordon model have been studied
in \cite{klimas2018oscillons, hahne2020scattering}.
Sphaleron collisions and decay in a model with a false vacuum were explored in
\cite{lima2021scattering, anco2025long}. 

Here we study the collision of Lorentz-boosted oscillons in
smooth, quartic scalar field theories, free of
discontinuities. We find a rich and intricate scattering structure and,
where possible, provide analytical insight into its underlying
mechanism. In particular, we investigate the important question whether
oscillon collisions can trigger false to true vacuum decay, thereby
initiating a first-order phase transition; this cannot be addressed in the
signum-Gordon model. We find that colliding oscillons with sufficient
energy to cross the sphaleron barrier can indeed do this, and in
certain cases pass very close to the sphaleron solution. 

The outline of the paper is as follows. In Sec.~\ref{sec:pos-mod}, we
introduce the normal class of field theories, with
positive quartic self-interaction. We review the small-amplitude,
Fodor oscillon solutions and discuss the sphaleron solution and the
sphaleron's perturbative spectrum. In Sec.~\ref{sec:neg-mod} we
consider the inverted class obtained through analytic continuation
of the field and parameters, which produces a negative quartic
interaction. Here also, there are oscillon and sphaleron solutions.
In Sec.~\ref{inner-force} we find an expression for the
force between two well-separated, Fodor oscillons. The numerical
simulations of oscillon and sphaleron collisions -- the heart of
the paper -- are presented in Sec.~\ref{results}. Finally, in
Sec.~\ref{sec:conc} we offer some conclusions and discuss possible
future work.

\section{Field Theories with Positive Quartic Term}
\label{sec:pos-mod}

\begin{figure}
    \centering
    \includegraphics[width=0.6\linewidth]{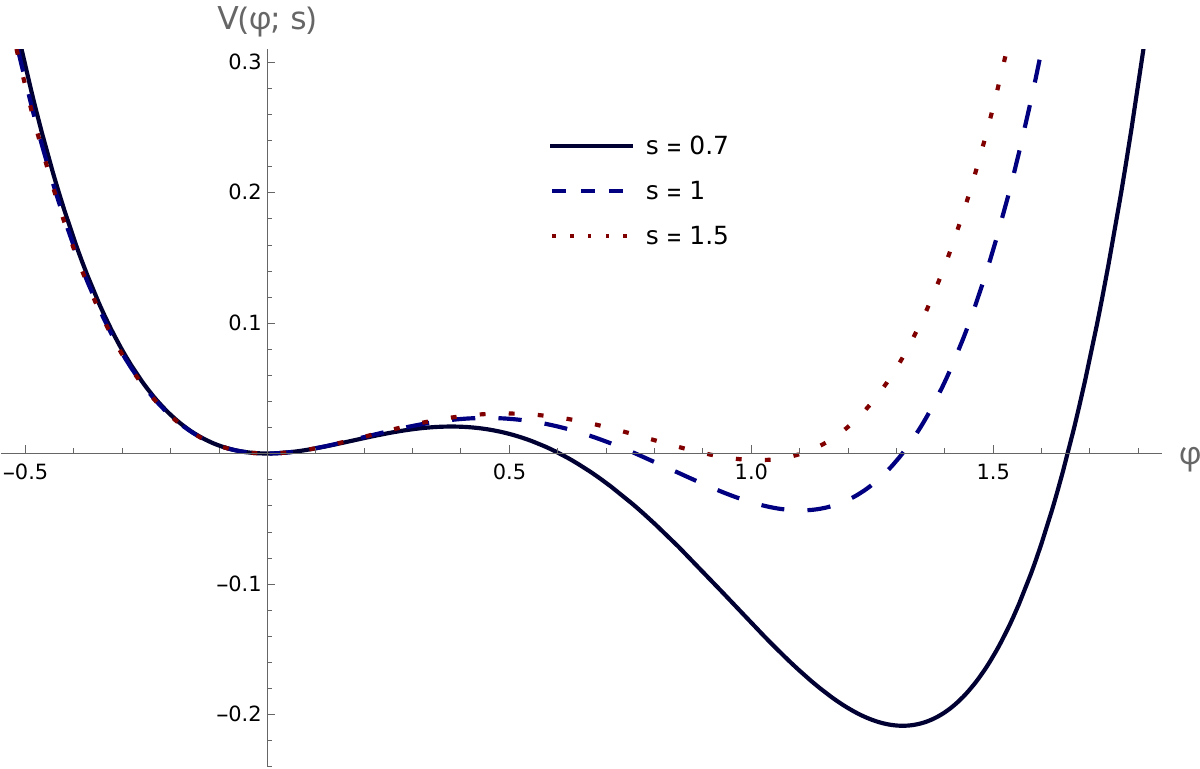}
   
    \caption{Potential in the normal model as a function of $\phi$ for
      several values of $s$.}
    \label{fig:pot1}
\end{figure}

Consider a real scalar field $\phi(x,t)$ in $(1+1)$-dimensions,
governed by the Lagrangian density
\begin{align}
\mathcal{L}=\frac 12 \partial_{\mu} \phi \, \partial^{\mu}\phi-V(\phi;s),
\end{align}
with potential \cite{avelar2008new}
\begin{align} \label{pot1}
V(\phi;s)=\frac {\phi^2}{2}\left(\phi-\tanh(s)\right)
\left(\phi-\coth(s)\right)=\frac{\phi^2}{2}
\left(1-2\alpha\phi + \phi^2\right).
\end{align}
In this normal model, the potential is bounded below. We have defined $\alpha = \coth(2s)$ where
$s$ is a positive control parameter, or equivalently $\alpha>1$. The
field equation is
\begin{align}
\label{fieldeqnorm}
\phi_{t t}-\phi_{x x}=-\phi(1+g_2 \phi+g_3\phi^2),
\end{align}
where $g_2=- 3 \alpha $ and $g_3= 2$. The
potential is zero at three points: $\phi=0$, which is a local
minimum (i.e. a false vacuum), $\phi=\tanh(s)$ and $\phi=\coth(s)$.
The global minimum of the potential is at
\begin{align}
\phi_{\rm{gmin}}=\frac 14 \left(3\alpha + \sqrt{9\alpha^2 - 8}\right),
\end{align}
and there is a local maximum at
\begin{align}
\phi_{\rm{lmax}}= \frac 14 \left(3\alpha - \sqrt{9\alpha^2 - 8}\right).
\end{align}
The potential for several values of $s$ is shown in Fig.~\ref{fig:pot1}.
As $s\to\infty$, the potential approaches that of the
standard $\phi^4$ model.

To construct small-amplitude oscillons, following Fodor \emph{et al.}
we introduce the rescaled space and time variables
\begin{align}
\zeta=\epsilon x, \quad \tau=\omega t,
\label{eq:fodor-freq}
\end{align}
where $\epsilon$ is small, and the Fodor frequency is
$\omega=\sqrt{1-\epsilon^2}$. The field equation becomes
\begin{align}
(1-\epsilon^2)\phi_{\tau \tau}-\epsilon^2\phi_{\zeta \zeta}
=-\phi(1+g_2 \phi+g_3\phi^2),
\end{align}
and we seek solutions of the form
\begin{align}
\label{eq:oscillon-exp}
\phi_{\rm F}(\zeta,\tau)=\sum_{k=1}^{\infty} \epsilon^k \phi_k(\zeta,\tau).
\end{align}
Substituting this expansion into the field equation yields the
recursive system
\begin{align}
\phi_{k,\tau\tau}
- \phi_{k-2,\tau\tau}
- \phi_{k-2,\zeta\zeta}
+ \phi_k
+ g_2\!\!\sum_{\substack{i+j=k \\ i,j\geq 1}}\!\phi_i\,\phi_j
+ g_3\!\!\sum_{\substack{i+j+\ell=k \\ i,j,\ell\geq 1}}\!\phi_i\,\phi_j\,\phi_\ell
=0,
\qquad k\geq 1 .
\end{align}
Solving these order by order, one finds the solutions 
\begin{align}
\phi_{1}(\zeta,\tau) &= p_{1}(\zeta)\,\cos(\tau), \\[6pt]
\phi_{2}(\zeta,\tau) &= -\tfrac{g_2}{6}\, p_{1}^{2}(\zeta)\,( 3-\cos(2\tau))\, , \\[6pt]
\phi_{3}(\zeta,\tau) &= p_{3}(\zeta)\,\cos(\tau) 
+ \tfrac{1}{72}\, p_{1}^{3}(\zeta)\,(4g_2^2-3\lambda)\, \cos(3\tau), 
\end{align}
up to the third order, where
\begin{align}
p_{1}(\zeta) &=\sqrt{\frac{2}{\lambda}}\,\sech(\zeta), \\[6pt]
p_{3}(\zeta)&=\frac{280 g_{2}^{4} - 132 g_{2}^{2} \lambda + 9 \lambda^{2}}
{81 \sqrt{2}\,\lambda^{5/2}}\sech(\zeta)
-\frac{ \left( 280 g_{2}^{4} + 192 g_{2}^{2} \lambda + 9 \lambda^{2} \right) }
{162 \sqrt{2}\,\lambda^{5/2}}\sech^{3}(\zeta).
\end{align}
The Fodor parameter $\lambda=\frac{5}{6}g_2^2-\frac{3}{4}g_3
=\frac{15}{2}\alpha^2-\frac{3}{2}$ is always positive here, as is
required for an oscillon to exist. In terms of $\alpha$, the solutions
up to the third order take the form
\begin{align}
\phi_{1}(\zeta,\tau)
  &=\frac{2}{\sqrt{3\,(5\alpha^{2}-1)}}\,\sech(\zeta)\,\cos(\tau), \\[6pt]
\phi_{2}(\zeta,\tau) &= \frac{2\alpha}{3\,(5\alpha^{2}-1)}\,\text{sech}^{2}(\zeta)\,( 3-\cos(2\tau))\, , \\[6pt]
\phi_{3}(\zeta,\tau) &= \left(\frac{1 + 78\alpha^{2} + 705\alpha^{4}}{9\sqrt{3}\,\bigl(5\alpha^{2}-1\bigr)^{5/2}}
\,\text{sech}(\zeta)
\;-\;
\frac{1 - 138\alpha^{2} + 1785\alpha^{4}}{18\sqrt{3}\,\bigl(5\alpha^{2}-1\bigr)^{5/2}}
\,\text{sech}^{3}(\zeta)\right)\,\cos(\tau) \nonumber
\\& \qquad + \frac{1+3\alpha^{2}}{6\sqrt{3}\,\bigl(5\alpha^{2}-1\bigr)^{3/2}}\,\text{sech}^{3}(\zeta)\, \cos(3\tau). 
\end{align}

\begin{figure}
    \centering
    \includegraphics[width=0.6\linewidth]{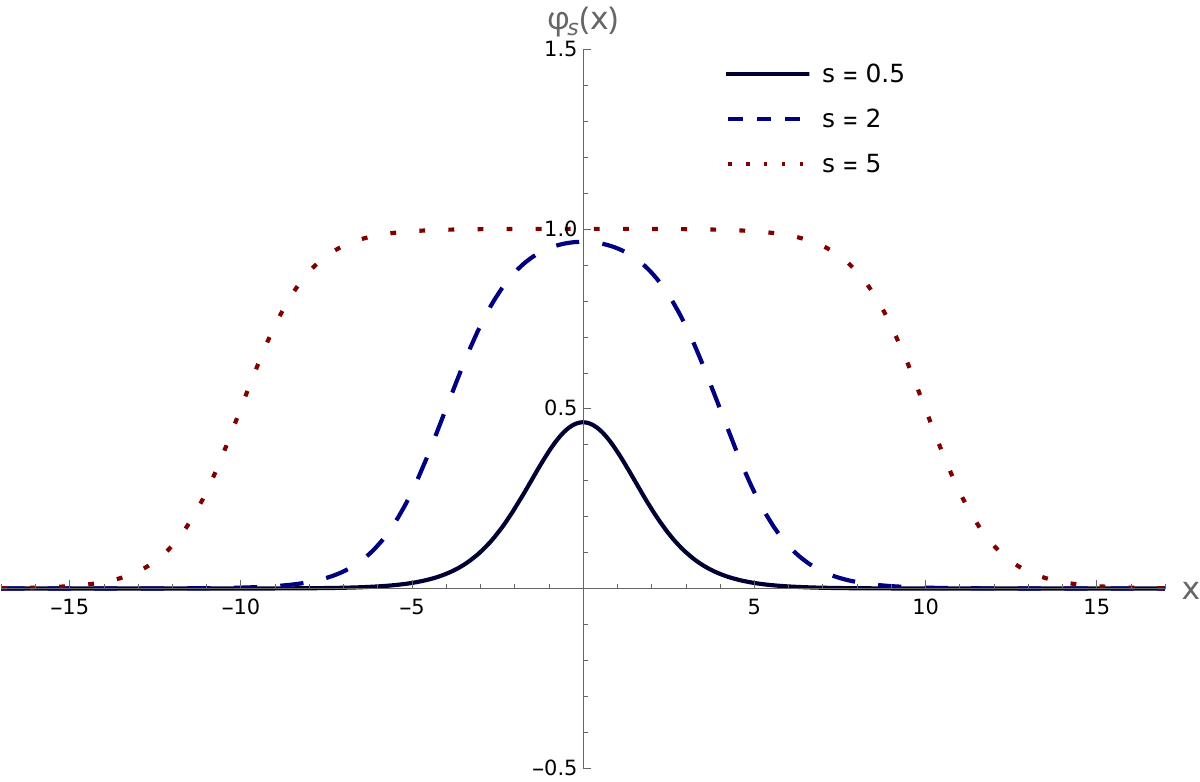}
   
    \caption{Sphaleron profiles $\phi_S(x)$ for several values of $s$.}
    \label{fig:Sphaleron1}
\end{figure}

\begin{figure}
    \centering
    \includegraphics[width=0.65\linewidth]{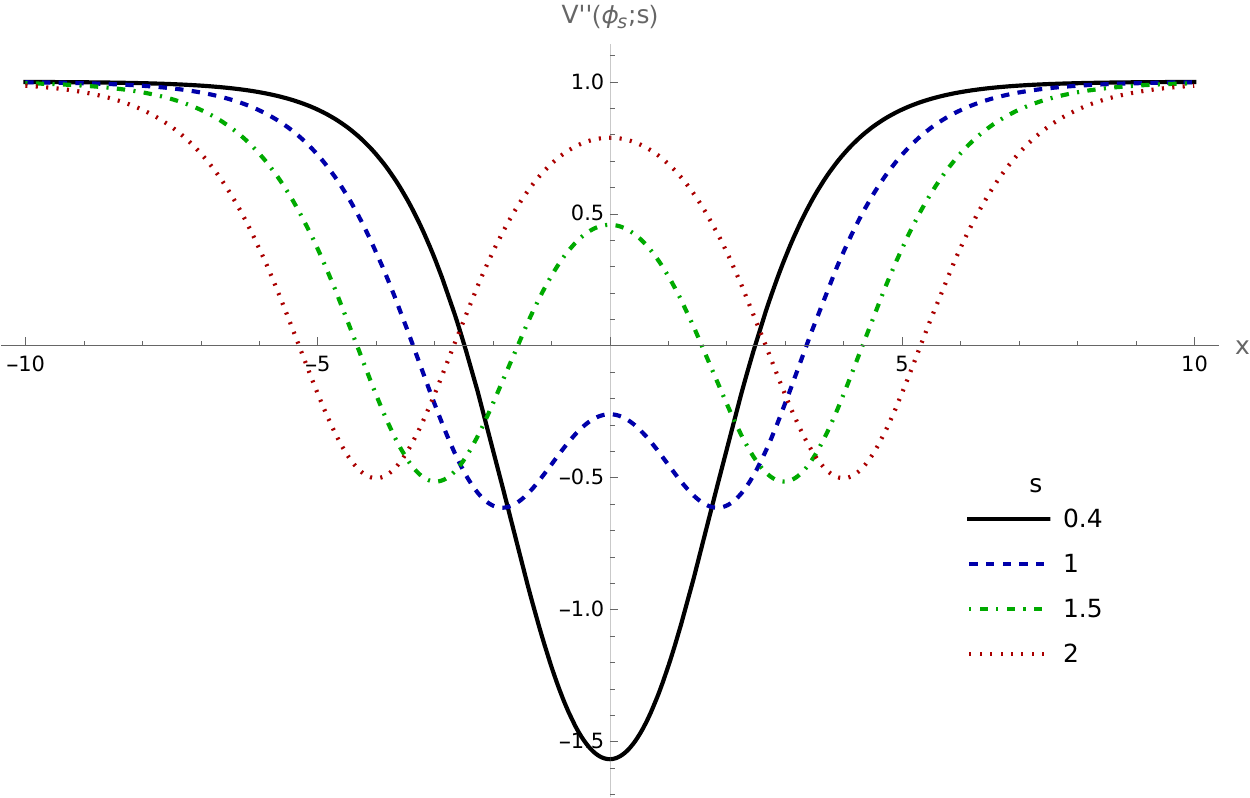}

    \caption{Stability potential $V''(\phi_S(x;s))$ as function of $x$.}
    \label{stability_pot1}
  \end{figure}

\begin{figure}
    \centering
    \includegraphics[width=0.6\linewidth]{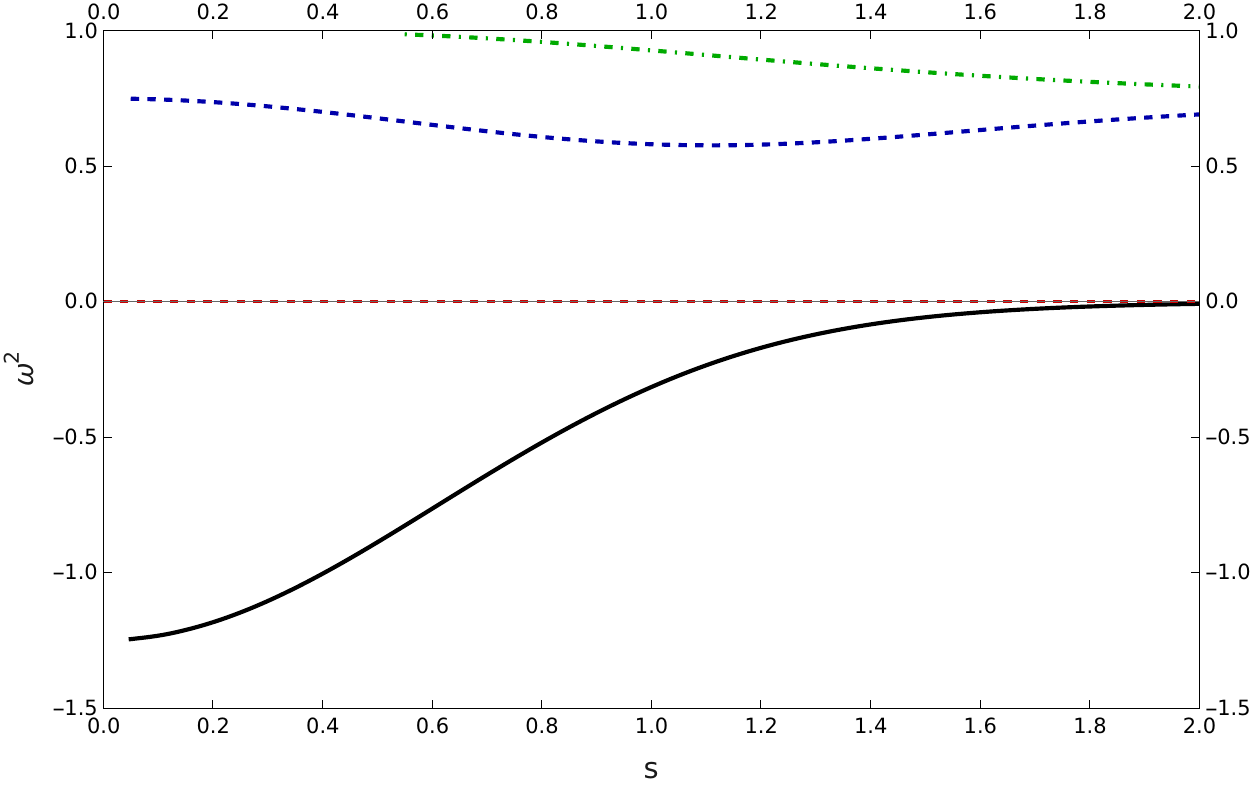}
   
    \caption{Small oscillation spectrum around the sphaleron
      $\phi_S$ as function of $s$.}
    \label{spectrum_sphaleron1}
\end{figure}

The field equation (\ref{fieldeqnorm}) also has an exact, static
sphaleron solution running from $\phi = 0$ to
$\phi = \tanh(s)$ and back, 
\begin{align}
\phi_S(x;s)=\frac 12 \left[\tanh\left(x/2+s\right)
-\tanh\left(x/2-s\right)\right]=\frac{\sinh(2s)}{\cosh(x)+\cosh(2s)}.
\label{eq:sphaleron}
\end{align}  
Sphaleron profiles are displayed in Fig.~\ref{fig:Sphaleron1}.
The stability potential $V''(\phi_S(x;s))$ is shown
in Fig.~\ref{stability_pot1}, and the resulting spectrum of small oscillations
in Fig.~\ref{spectrum_sphaleron1}. The spectrum
contains a single unstable mode with a negative squared frequency,
characteristic of a sphaleron. The mass of the sphaleron is given by 
\begin{align}
    M_S(s)=\coth^2{(2s)}-2s \, \coth{(2s)} \,\textrm{csch}^2{(2s)}-\frac23.
\end{align}

\section{Field Theories with Negative Quartic Term}
\label{sec:neg-mod}

The inverted model with negative quartic coupling 
($g_3<0$) is obtained by the analytic continuation
\[
\phi \;\rightarrow\; i\psi .
\]
To keep the kinetic energy positive, we simultaneously change the sign
of the Lagrangian density, $\mathcal{L}\rightarrow -\mathcal{L}$, 
and additionally take the parameter $s$ to be pure imaginary,
\[
s \;\rightarrow\; i r, 
\qquad r\in\mathbb{R}.
\]
With these definitions, the potential becomes
\begin{align} \label{pot2}
U(\psi;r)
&= -V(i\psi;ir)
= -\frac{\psi^2}{2}
   \left(\psi-\tan(r)\right)
   \left(\psi+\cot(r)\right)= \frac{\psi^2}{2}
   \left(1-2\beta \psi-\psi^2\right),
\end{align}
where $\beta=(\cot(r)-\tan(r))/2=\cot{(2r)}$. $\psi = 0$ is again a
false vacuum, but the potential is unbounded below, so there is no
true vacuum. The corresponding field equation is
\begin{align}
\psi_{t t}-\psi_{x x}=-\psi(1+g_2 \psi+g_3\psi^2),
\end{align}
with couplings $g_2=-3\beta$ and $g_3=-2$.

\begin{figure}
    \centering
    \includegraphics[width=0.6\linewidth]{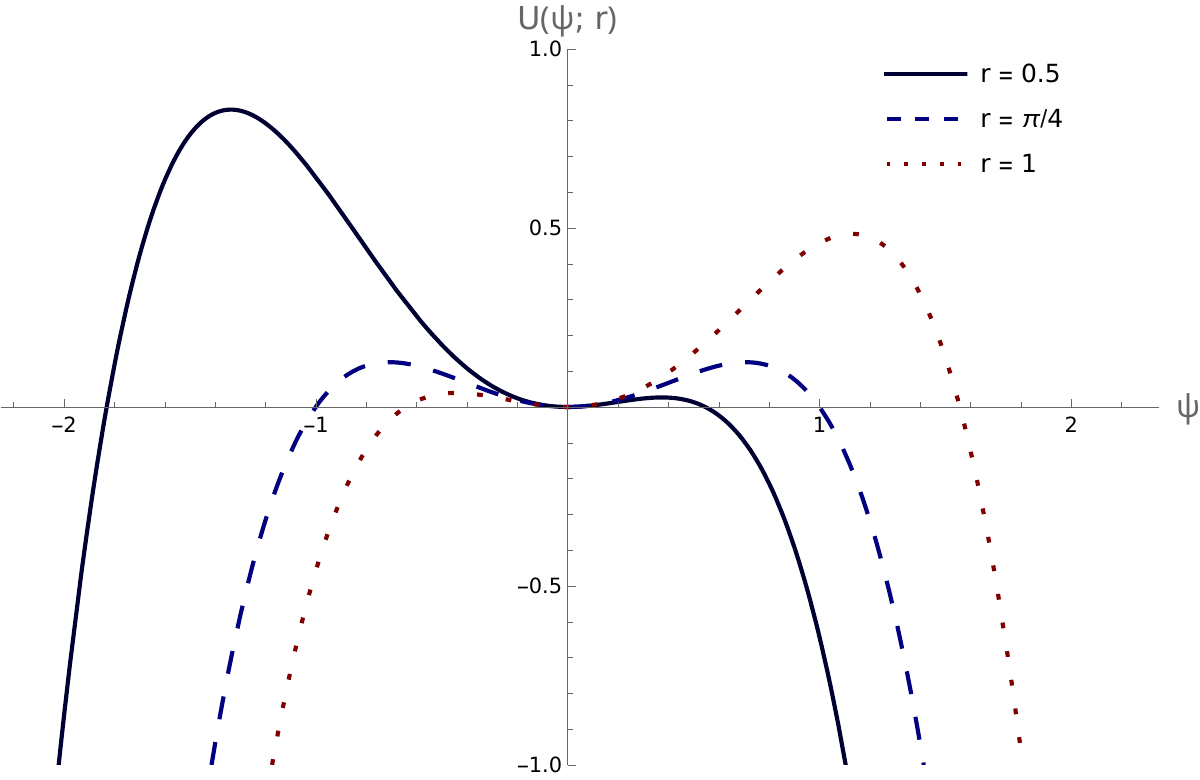}
   
    \caption{Potential in the inverted model as a function of $\psi$
      for several values of $r$.}
    \label{fig:pot2}
\end{figure}

For $0 < r < \pi/2$, the potential is zero at three points:
$\psi=0$, a local minimum, $\psi=\tan(r)$ and
$\psi=-\cot(r)$. In the range $0<r<\pi/4$, the potential has a local
maximum at
\begin{align}
\psi_{\rm{lmax}}= \frac 14 \left(-3\beta + \sqrt{9\beta^2 + 8}\right),
\end{align}
and a global maximum at
\begin{align}
\psi_{\rm{gmax}}= \frac 14 \left(-3\beta - \sqrt{9\beta^2 + 8}\right).
\end{align}
For $\pi/4<r<\pi/2$, the roles of these maxima are reversed.  
At the symmetric point $r=\pi/4$, the two coincide and the
potential becomes symmetric about $\psi=0$. The potential $U(\psi;r)$
for several values of $r$ is shown in Fig.~\ref{fig:pot2}.
The parameter $\lambda=\frac{15}{2}\beta^2+\frac{3}{2}$
is again always positive, guaranteeing the existence of
oscillons around $\psi = 0$.

Using the same rescaled variables $(\zeta,\tau)$ and small-amplitude
expansion as in the normal model, the oscillon solutions $\psi_{\rm F}$ up
to third order take the form
\begin{align}
\psi_{1}(\zeta,\tau) &= \frac{2}{\sqrt{3\,(5\beta^{2}+1)}}\, \text{sech}(\zeta) \,\cos(\tau), \\[6pt]
\psi_{2}(\zeta,\tau) &= \frac{2\beta}{3\,(5\beta^{2}+1)}\,\text{sech}^{2}(\zeta)\,( 3-\cos(2\tau))\, , \\[6pt]
\psi_{3}(\zeta,\tau) &= \left(\frac{1 - 78\beta^{2} + 705\beta^{4}}{9\sqrt{3}\,\bigl(5\beta^{2}+1\bigr)^{5/2}}
\,\text{sech}(\zeta)
\;-\;
\frac{1 + 138\beta^{2} + 1785\beta^{4}}{18\sqrt{3}\,\bigl(5\beta^{2}+1\bigr)^{5/2}}
\,\text{sech}^{3}(\zeta)\right)\,\cos(\tau) \nonumber
\\& \qquad + \frac{-1+3\beta^{2}}{6\sqrt{3}\,\bigl(5\beta^{2}+1\bigr)^{3/2}}\,\text{sech}^{3}(\zeta)\, \cos(3\tau).
\end{align}

\begin{figure}
    \centering
\includegraphics[width=0.6\linewidth]{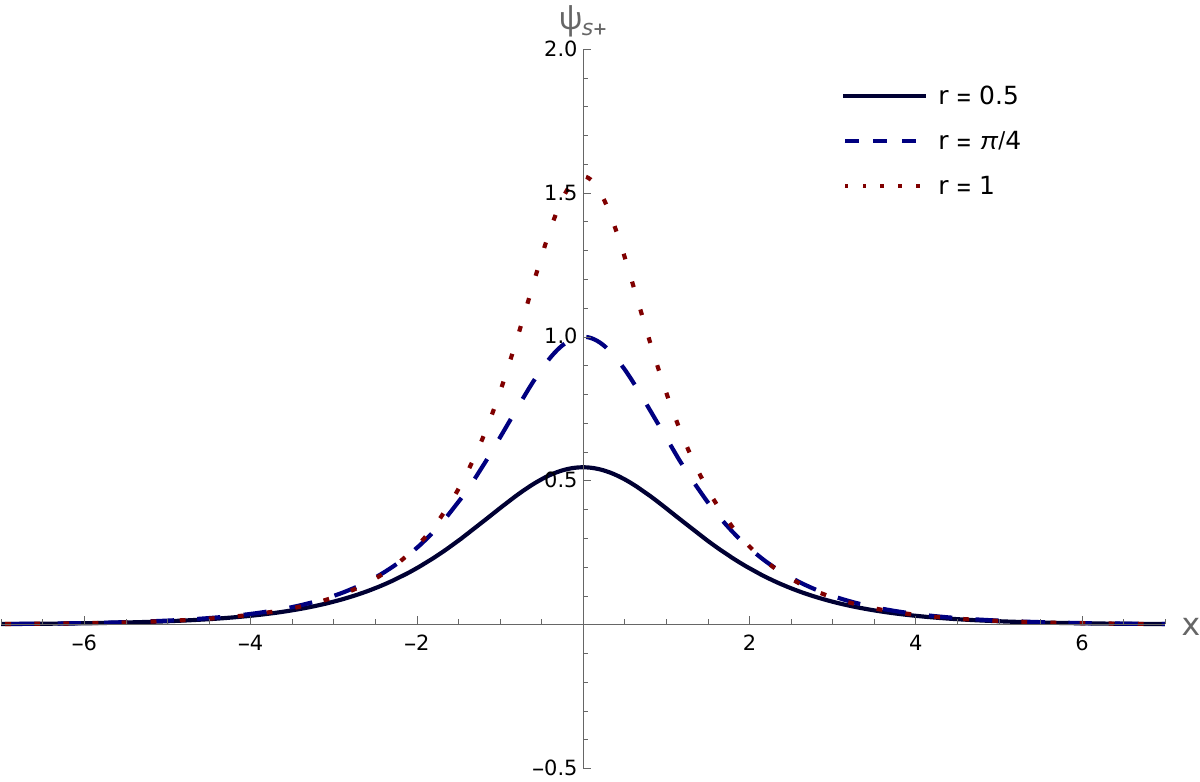}
   
\caption{Sphaleron profiles $\psi_{S,+}(x)$ for several values of $r$.}
    \label{fig:sphaleron2p}
\end{figure}

The inverted model also admits two distinct sphalerons,
\begin{align}
\psi_{S,+}(x;r)&=-\frac {i}{2}\left[\tanh\left(x/2+i r\right)-\tanh\left(x/2-i r\right)\right]=\frac{\sin(2r)}{\cosh(x)+\cos(2r)},\\
\psi_{S,-}(x;r)&=-\frac {i}{2}\left[\coth\left(x/2+i r\right)-\coth\left(x/2-i r\right)\right]=-\frac{\sin(2r)}{\cosh(x)-\cos(2r)},
\end{align}
related by
\begin{align}
\psi_{S,-}\!\left(x;\tfrac{\pi}{2}-r\right) = -\,\psi_{S,+}(x;r).
\end{align}
Sphaleron profiles $\psi_{S,+}$ are displayed in
Fig.~\ref{fig:sphaleron2p}.

\begin{figure}
    \centering
    \includegraphics[width=0.65\linewidth]{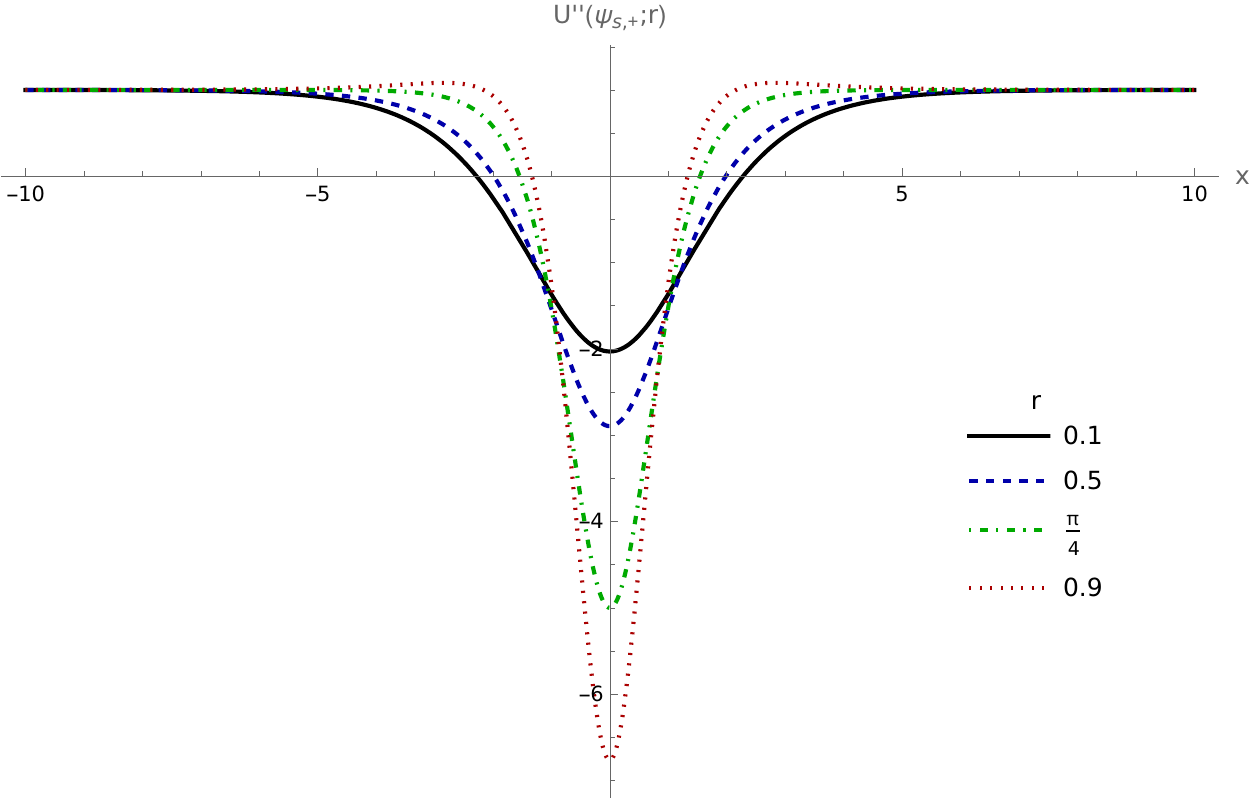}
   
    \caption{Stability potential $U''(\psi_{S,+};r)$ as function of $x$.}
    \label{stability_pot2p}
\end{figure}

\begin{figure}
    \centering
    \includegraphics[width=0.6\linewidth]{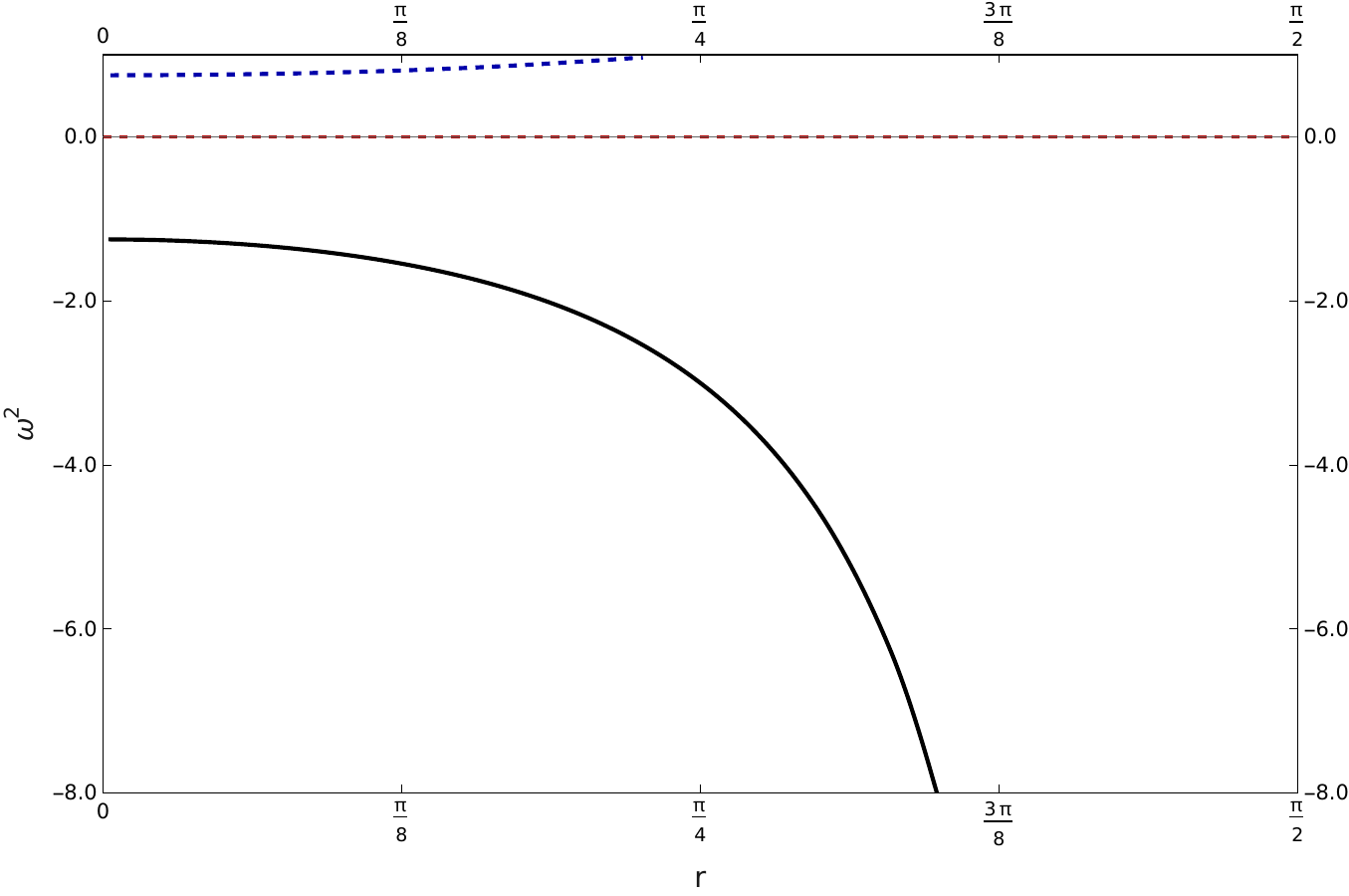}
   
    \caption{Small oscillation spectrum around the sphaleron
      $\psi_{S,+}$ as function of $r$.}
    \label{spectrum_sphaleron2p}
\end{figure}

The stability potential for $\psi_{S,+}$ is
\begin{align}
  U''(\psi_{S,+};r) = 1 - 6 \cot(2r)\,\psi_{S,+}(x;r) - 6\,\psi^2_{S,+}(x;r),
\end{align}
while $U''(\psi_{S,-};r)$ is obtained by the
replacement $r\to\pi/2-r$. $U''(\psi_{S,+};r)$ is displayed in
Fig.~\ref{stability_pot2p}, and the spectrum of small oscillations
in Fig.~\ref{spectrum_sphaleron2p}. For all $r$, the sphaleron has
a single unstable mode. The masses of the sphalerons are given by 
\begin{align}
   & M_{S,+}(r)=\cot^2{(2r)}-2r \, \cot{(2r)} \,\textrm{csc}^2{(2r)}+\frac23,\nonumber\\
   & M_{S,-}(r)=M_{S,+}(\pi/2-r).
\end{align}

\section{The Inter-Oscillon Force}
\label{inner-force}

The interaction between two well-separated, small-amplitude
oscillons can be derived using a simplified ansatz based on the
leading term in the Fodor expansion,
\begin{align}
\phi(x,t)
&= A \epsilon_1 \sech\big(\epsilon_1(x+a)\big)\cos\theta_{1}(t) + A \epsilon_2 \sech\!\big(\epsilon_2(x- a)\big)\cos\theta_{2}(t),
\end{align}
where $A=\sqrt{2/\lambda}$, $\theta_1=\omega_1 t+\delta_1$, 
$\theta_2=\omega_2 t+\delta_2$, 
and the oscillons are centered at $\pm a$. The parameters $\epsilon_i$
determine the oscillons' amplitudes and widths, and their frequencies
$\omega_i=\sqrt{1-\epsilon_i^2}$.

The interaction force is obtained from the rate of change of the total
momentum
\[
P=\int_{-\infty}^{X}T^{01}\,dx,
\]
contained in the half-line to the left of an intermediate point $X$.
Using the local conservation law for the stress-energy tensor,
$\partial_\mu T^{\mu\nu}=0$, one finds
\begin{align}
F=\frac{dP}{dt}
=\left[-\frac12 (\partial_t \phi)^2
       -\frac12 (\partial_x \phi)^2
       +V(\phi)\right]_{-\infty}^{X}.
\end{align}
The second equality follows from integrating the conservation equation
over the half-line, which relates the time derivative of the momentum to
the pressure evaluated at the boundary. The contribution from
$x=-\infty$ vanishes. For large $a$, and keeping
only the leading-order overlap between the exponential tails, the
force becomes
\begin{align}
F(a) \simeq
4A^2 \epsilon_1 \epsilon_2 \,
e^{-2a\bar{\epsilon}} e^{-X\Delta\epsilon}
\Big[(1+\epsilon_1 \epsilon_2)\cos\theta_1\cos\theta_2
-\omega_1\omega_2 \sin\theta_1\sin\theta_2\Big],
\end{align}
where $\bar \epsilon=(\epsilon_1+\epsilon_2)/2$ and
$\Delta\epsilon=\epsilon_1-\epsilon_2$. Different choices of $X$
correspond to different conventions for splitting the system. 
At leading order in the tail overlaps, however,  the physically relevant information is encoded in the 
relative phase dependence and in the universal exponential 
factor $e^{-2a\bar{\epsilon}}$. A convenient and symmetric choice is
$X = X_\ast$, such that the time-averaged magnitudes of the two tails coincide. This gives
\begin{equation}
X_\ast=\frac{a(\epsilon_2-\epsilon_1)+\ln(\epsilon_1/\epsilon_2)}
{\epsilon_1+\epsilon_2}.
\end{equation}
An alternative choice is $X=X_{\rm CM}$, the center of mass
of the configuration. The energy (mass) of a single oscillon with
parameter $\epsilon$ and phase $\theta(t)$ can be decomposed as
\begin{equation}
E(t)=E_{\rm kin}+E_{\rm grad}+E_{V_2}+E_{V_3}+E_{V_4},
\end{equation}
with
\begin{align}
E_{\rm kin} &= A^2\epsilon\,\omega^2\sin^2\theta, \quad
E_{\rm grad} = \frac{A^2\epsilon^3}{3}\cos^2\theta, \nonumber\\
E_{V_2} &= A^2\epsilon\cos^2\theta, \quad
E_{V_3} = \frac{\pi g_2}{6}A^3\epsilon^2\cos^3\theta, \quad
E_{V_4} = \frac{g_3 A^4\epsilon^3}{3}\cos^4\theta.
\end{align}
The time-averaged mass becomes
\begin{align}
M(\epsilon)
=
A^2 \epsilon
-\frac{A^2 \epsilon^3}{3}
+\frac{g_3 A^4 \epsilon^3}{8}.
\end{align}
The two-oscillon center of mass is then
\begin{align}
X_{\rm CM}
= a\,\frac{M(\epsilon_2)-M(\epsilon_1)}{M(\epsilon_1)+M(\epsilon_2)}.
\end{align}
Clearly, $X_{\rm CM}$ does not coincide with
$X_\ast$, although both reduce to $X=0$ in the symmetric case
$\epsilon_1=\epsilon_2$. $X_\ast$ seems the better choice, because the
force formula comes from the oscillon interaction region
where the exponential tails overlap, whereas $X_{\rm CM}$
is defined globally and is controlled mainly by the oscillon cores,
not the overlap region.

Using the trigonometric identities
\begin{align}
\sin\theta_1 \sin\theta_2
&=\tfrac12\!\left[\cos(\theta_1-\theta_2) -\cos(\theta_1+\theta_2)\right]\nonumber\\
\cos\theta_1 \cos\theta_2
&=\tfrac12\!\left[\cos(\theta_1-\theta_2)+\cos(\theta_1+\theta_2)\right],
\end{align}
and defining
\begin{align}
\Delta\omega=\omega_1-\omega_2,
\quad
\delta=\delta_1-\delta_2,
\quad
\Omega=\omega_1+\omega_2,
\quad
\tilde \delta=\delta_1+\delta_2,
\end{align}
the force can be rewritten as
\begin{align}
F(a)
& \simeq
2 A^2 \epsilon_1 \epsilon_2 \,
e^{-2a\bar{\epsilon}} e^{-X_\ast\Delta\epsilon}
\Big[
\big((1+\epsilon_1 \epsilon_2)-\omega_1\omega_2\big)
\cos(\Delta\omega \, t+\delta)
\nonumber\\
&+
\big((1+\epsilon_1 \epsilon_2)+\omega_1\omega_2\big)
\cos(\Omega t+\tilde \delta)
\Big].
\end{align}
For $\epsilon_1=\epsilon_2=\epsilon$, the force becomes
\begin{align}
\label{force-epsequal}
F(a)
\simeq
4A^2\epsilon^2e^{-2a\epsilon}
\Big[
\epsilon^2\cos\delta
+
\cos(2\omega t+\tilde\delta)
\Big].
\end{align}
To understand how the two terms affect the motion of the oscillons, we
treat $a$ as a collective coordinate and write
\[
M\ddot a
=
F_0(a)
+
F_{\rm osc}(a)\cos(2\omega t+\tilde\delta),
\]
where
\[
F_0(a)
=
4A^2\epsilon^4e^{-2\epsilon a}\cos\delta,
\qquad
F_{\rm osc}(a)
=
4A^2\epsilon^2e^{-2\epsilon a}.
\]
For well-separated oscillons and over the time interval in which the
separation changes slightly, we approximate
\[
F_0(a)\simeq F_0(a_0),
\qquad
F_{\rm osc}(a)\simeq F_{\rm osc}(a_0),
\]
where $a_0$ is the initial separation. Integrating twice with
$a(0)=a_0$ and $\dot a(0)=0$ gives
\[
a(t)
=
a_0
+
\frac{F_0(a_0)}{2M}t^2
+
\frac{F_{\rm osc}(a_0)}{4M\omega^2}
\Big[
\cos\tilde\delta
-
\cos(2\omega t+\tilde\delta)
\Big]
-
\frac{F_{\rm osc}(a_0)}{2M\omega}
\sin\tilde\delta\,t.
\]
The constant force therefore produces a displacement proportional to
$t^2$, whereas the oscillatory force produces a bounded oscillation and,
for $\sin\tilde\delta\neq0$, an additional linear drift. Since
$F_0\sim\epsilon^4$ and $F_{\rm osc}\sim\epsilon^2$, the bounded
oscillatory contribution remains of order $\epsilon^2$, while the
constant-force displacement grows as
\[
\Delta a_0\sim\epsilon^4t^2.
\]
In the absence of the drift term, the two become comparable when
\[
\epsilon^2
\sim
\epsilon^4t^2,
\qquad
t\sim\frac1\epsilon.
\]
Thus, when $\sin\tilde\delta=0$, the constant force dominates the motion
after times of order $1/\epsilon$. If $\sin\tilde\delta\neq0$, the
oscillatory force also generates a linear drift,
$\Delta a_{\rm drift}\sim\epsilon^2t$, which becomes comparable to the
quadratic displacement only after
\[
\epsilon^2t
\sim
\epsilon^4t^2,
\qquad
t\sim\frac1{\epsilon^2}.
\]

We now compare these predictions with numerical simulations for
$\epsilon=0.15$ and three representative values of the relative phase:
$\delta=0$, $\pi/2$, and $\pi$. For $\delta=0$ and $\delta=\pi$,
$\sin\tilde\delta=0$, so the oscillatory contribution 
becomes negligible compared with the quadratic displacement after
$t\sim1/\epsilon\approx7$. In this regime, the force is well
approximated by
\begin{align}
\label{force-epsequal-final}
F(a)
\simeq
4A^2\epsilon^4\cos\delta\,e^{-2a\epsilon},
\end{align}
which is attractive for $\delta=0$ and repulsive for $\delta=\pi$.

\begin{figure}
    \centering
    \includegraphics[width=0.7\linewidth]{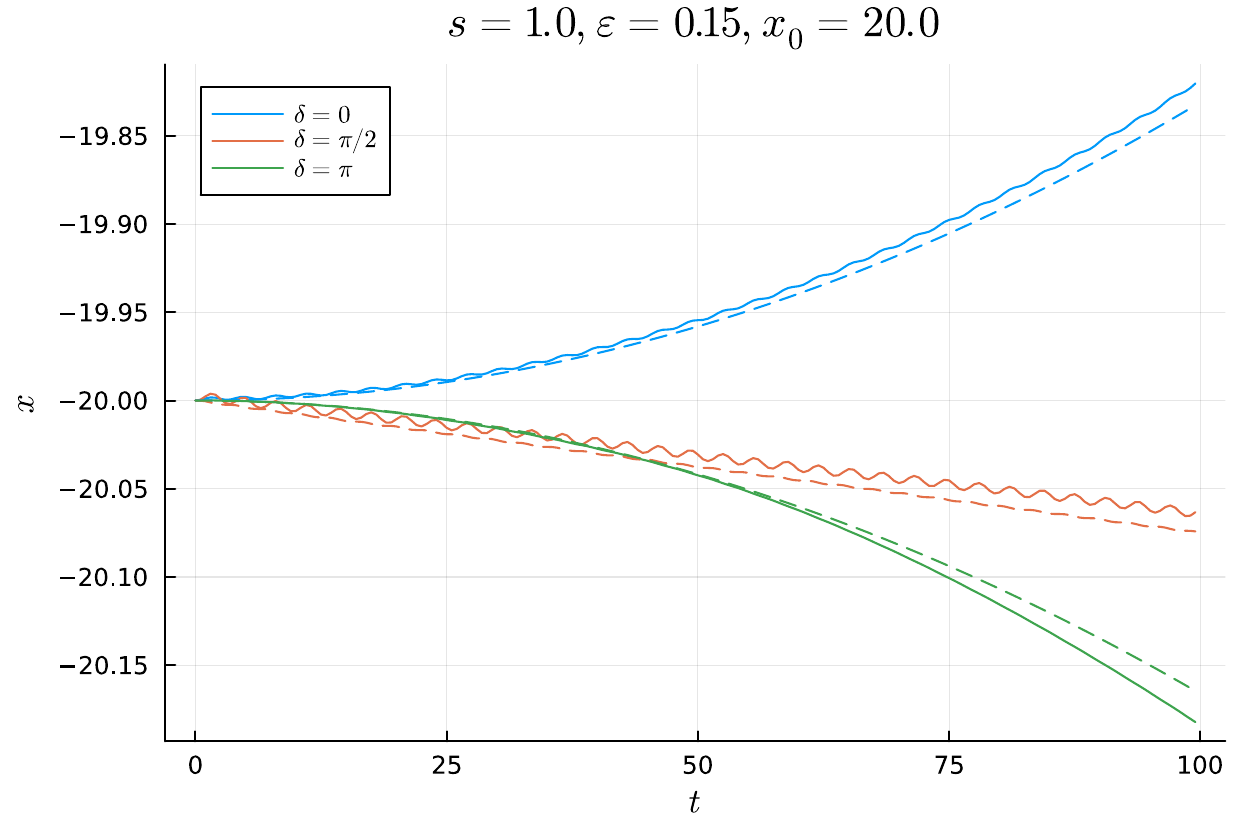}
    \caption{Trajectory of two oscillons. Solid curves show the result of the simulations, and the dashed ones show the contribution of the first term in (\ref{force-epsequal}) for $\delta=0$ and $\delta=\pi$, and the second term in (\ref{force-epsequal}) for $\delta=\pi/2$.}
    \label{fig:force}
\end{figure}

For $\delta=\pi/2$, the constant term vanishes identically, while
$\sin\tilde\delta=1$, giving the maximum possible drift. In this
case, the analytical prediction must be obtained from the oscillatory
term in Eq.~(\ref{force-epsequal}). As shown in
Fig.~\ref{fig:force}, the analytical and numerical trajectories agree
well in all three cases. The remaining small differences are partly due
to the fact that the numerical simulations are initialized using the
first three terms of the Fodor expansion, whereas the analytical force
is derived from the leading-order term alone.

\section{Numerical Simulations of Collisions}
\label{results}

In this section, we present the results of numerical oscillon and sphaleron
collisions. The field equation is integrated using the following numerical
algorithm. The partial derivatives are computed using a Fourier
spectral method; then, the resulting equations are integrated in time via a
fourth-order Runge-Kutta with step size control. Periodic spatial
boundary conditions at $x=\pm L$ with $L=200$ are employed, and to avoid
returning radiation, we add damping near these boundaries. The damping
term is smooth and has compact support.

\subsection{Single oscillon and sphaleron}

Here, for $s=1$, we illustrate the evolution of a Fodor oscillon for a
selection of fairly small amplitudes $\epsilon$, with
the initial condition given by eq.~\eqref{eq:oscillon-exp} truncated
at third order, and also the decay of a sphaleron \eqref{eq:sphaleron}.
To ensure the sphaleron evolves into an excited oscillon,
we give it a ``kick'' by taking as initial configuration an exact
sphaleron with shifted parameter $s+10^{-6}$; this brings the
kink-antikink pair composing the sphaleron closer together.

\begin{figure}
    \centering
    \includegraphics[width=0.8\linewidth]{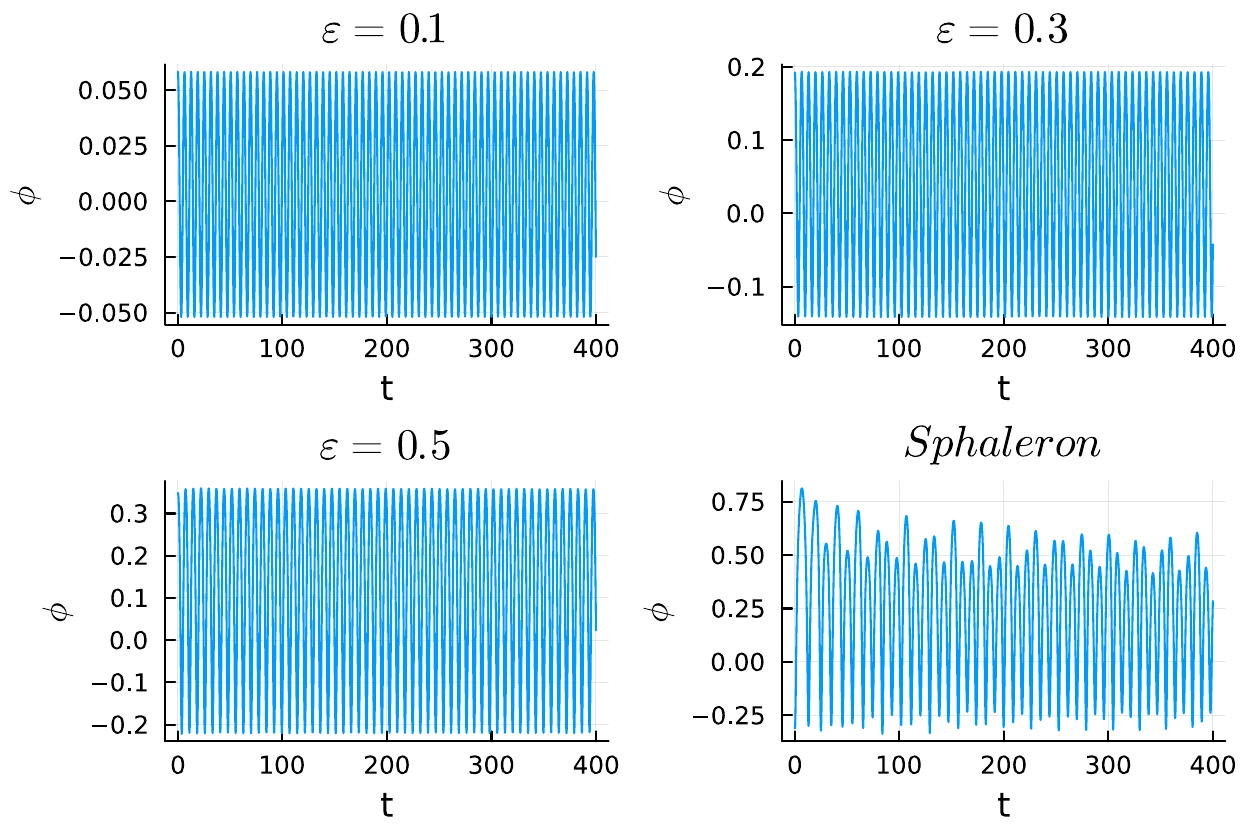}
    \caption{Time evolution of the field at the center of Fodor
      oscillons and a kicked sphaleron.}
    \label{fig:time-ev}
\end{figure}

\begin{figure}
    \centering
    \includegraphics[width=0.98\linewidth]{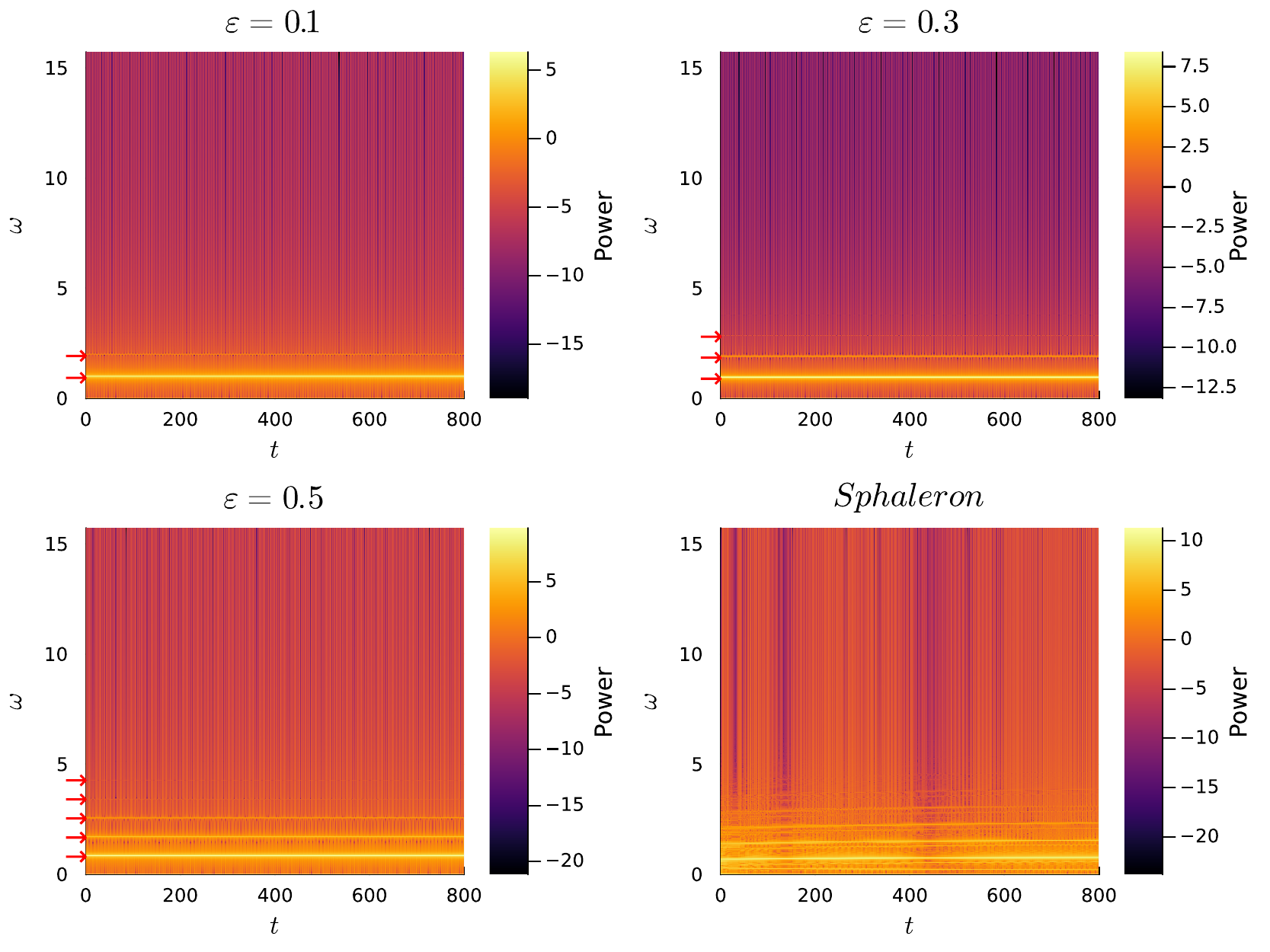}
    \caption{Power spectra of Fodor oscillons and a kicked sphaleron,
      as functions of $\omega$ and $t$. The Fodor frequencies are
      marked with red arrows.}
    \label{fig:PS}
\end{figure}

The time evolution of the field at the center of mass is shown in
Fig.~\ref{fig:time-ev}. The Fodor oscillons behave regularly, with nearly
periodic evolution, while the kicked sphaleron emits a transient burst of
radiation before settling into a large-amplitude oscillon having
erratic behaviour. We have performed a Fourier transform on
$\Delta t=200$ time intervals with sampling frequency $f=5$ per time
unit. The result for many initial times $t$ is shown in Fig.~\ref{fig:PS}.
Interestingly, the oscillon spectra remain nearly constant in time.
The spectral lines occur at the Fodor frequency $\omega=\sqrt{1-\epsilon^2}$
and its harmonics, marked by red arrows in the figure.

\begin{figure}
    \centering
    \includegraphics[width=0.73\linewidth]{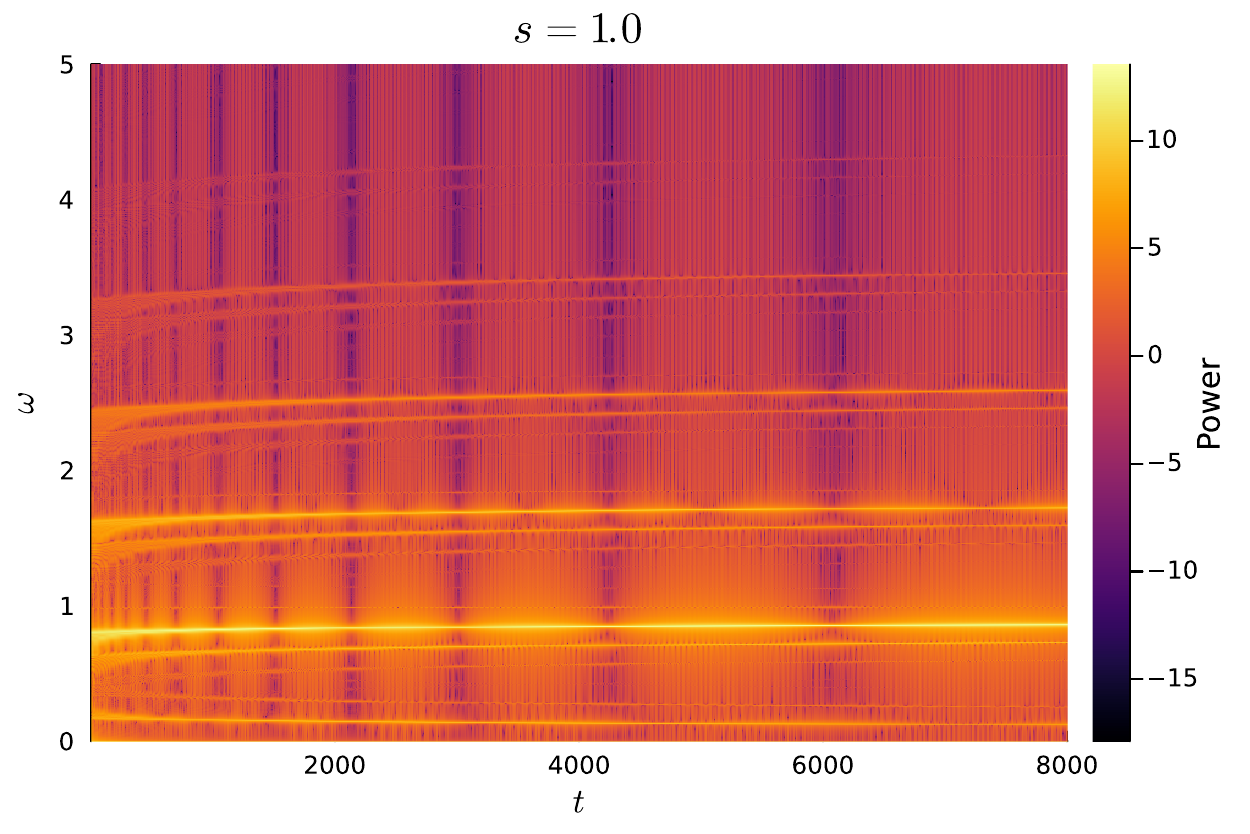}
    \caption{Power spectrum of a kicked sphaleron as function of
      $\omega$ and $t$.}
    \label{fig:FS}
\end{figure}

By contrast, the power spectrum of the kicked sphaleron varies in
time, as shown in Fig. \ref{fig:FS}, and has a fine structure on
top of the Fodor frequencies.  This
indicates that it is an excited oscillon. In fact, it is known from
ref.~\cite{blaschke2024amplitude} that excited oscillons exhibit amplitude
modulations, leading to the observed fine spectral structure, and they
can be interpreted as a bound pair of two oscillons.

\subsection{Oscillon collisions}

\begin{figure}
    \centering
    \includegraphics[width=0.32\linewidth]{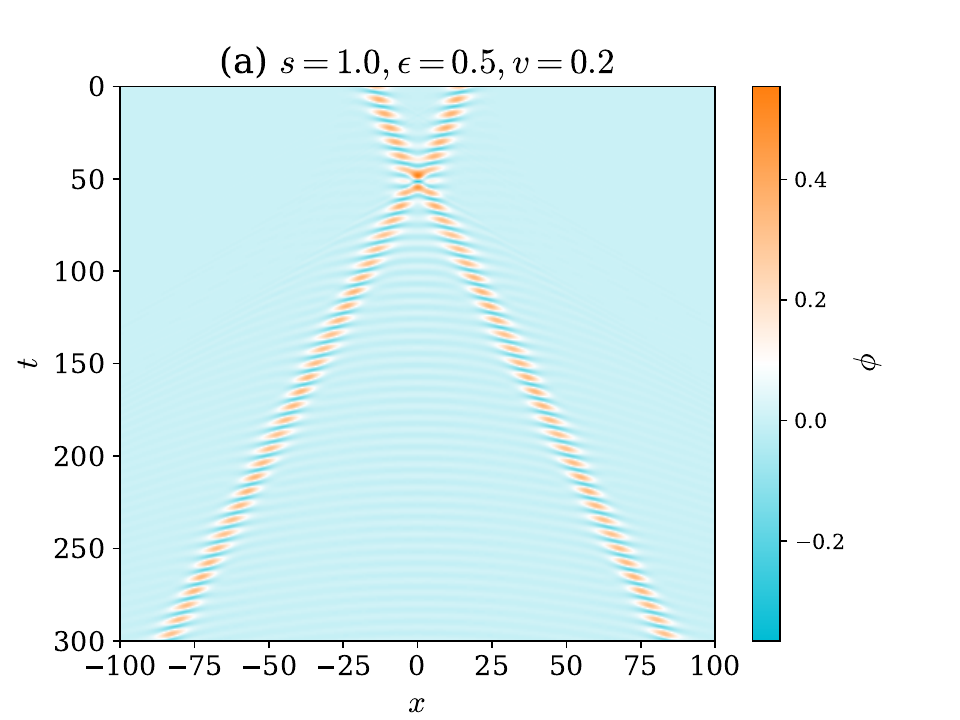}
    \includegraphics[width=0.32\linewidth]{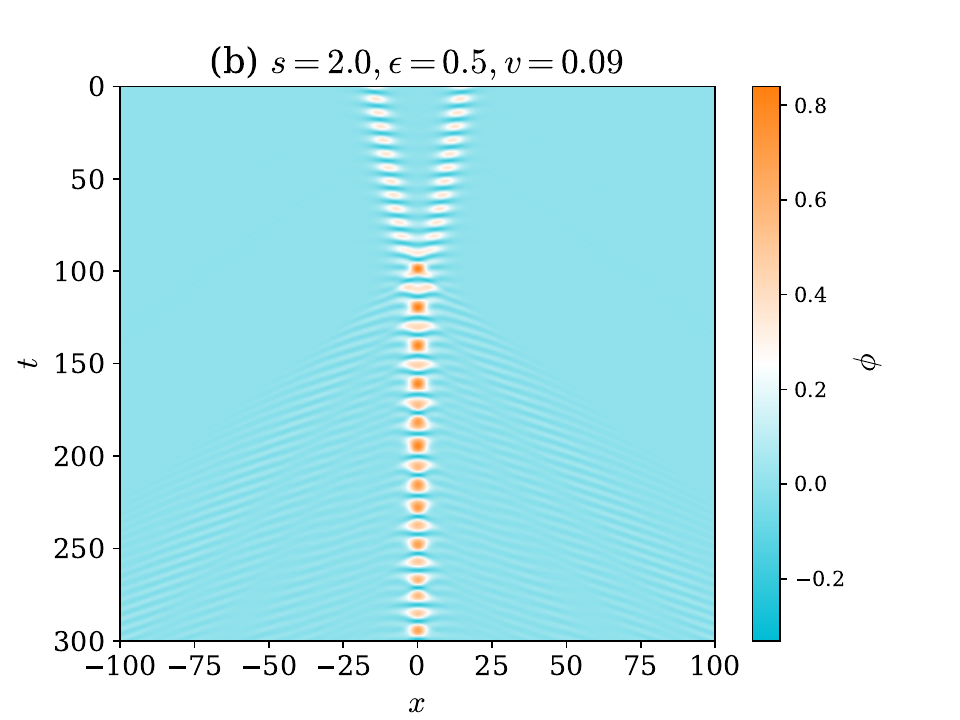}
    \includegraphics[width=0.32\linewidth]{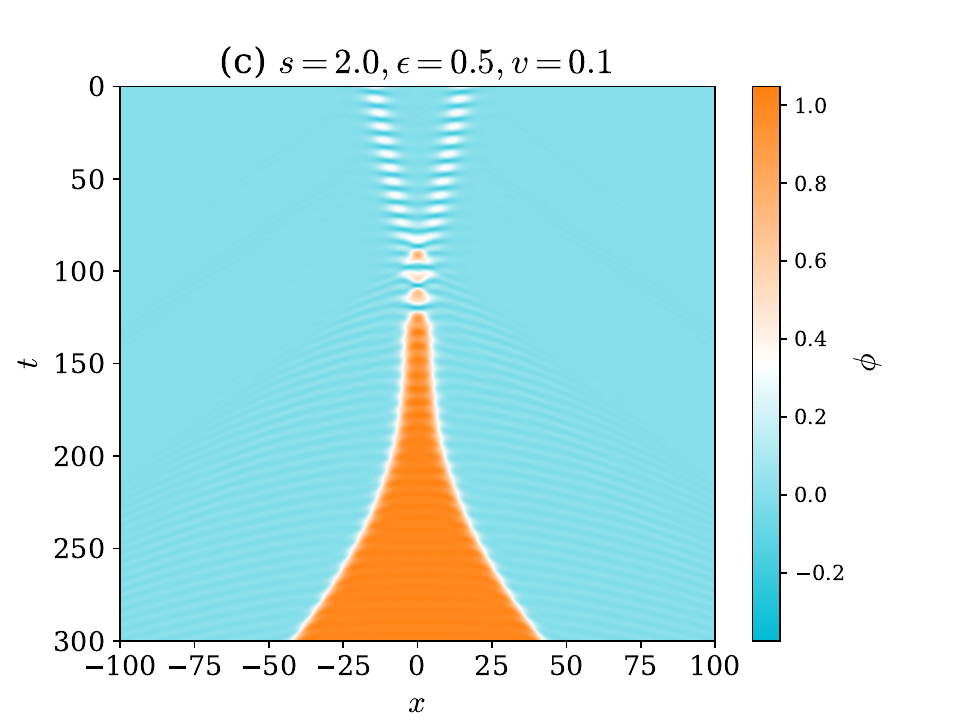}
    \includegraphics[width=0.32\linewidth]{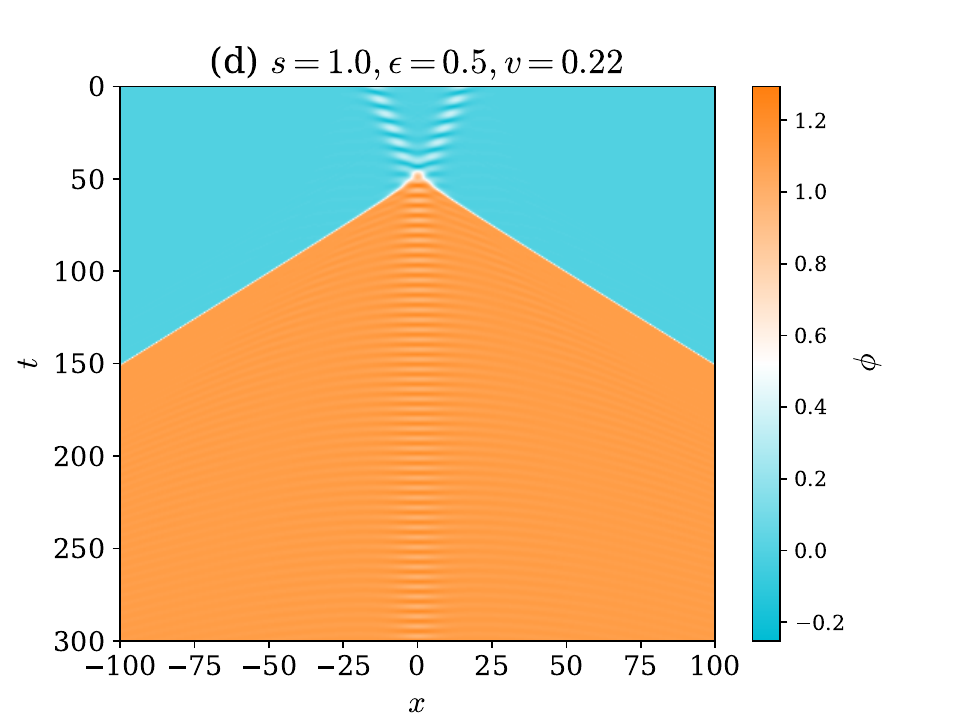}
    \includegraphics[width=0.32\linewidth]{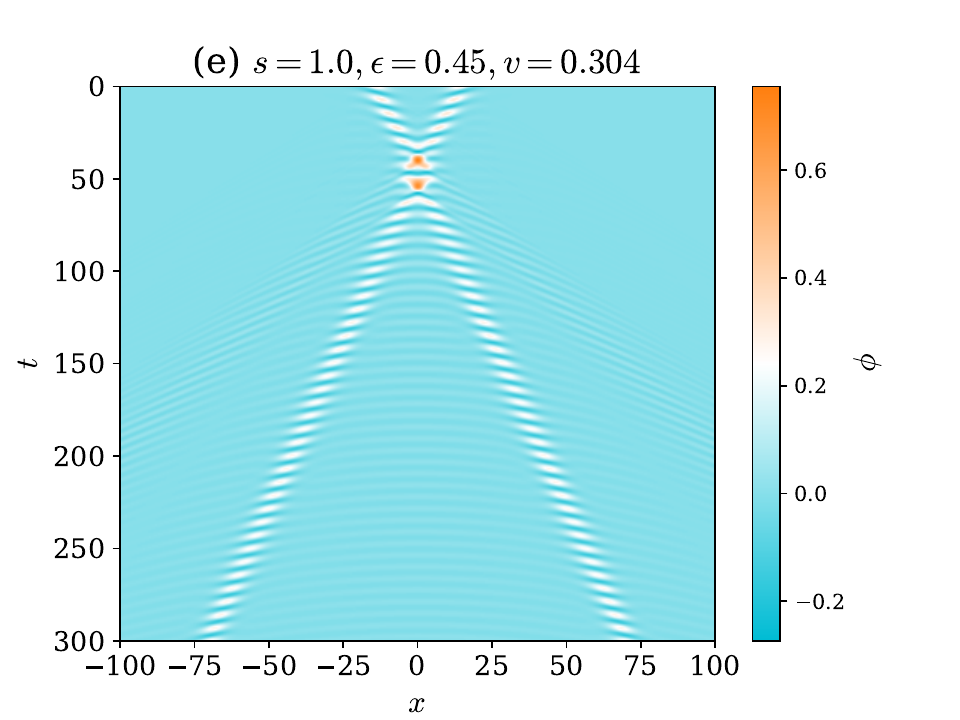}
    \includegraphics[width=0.32\linewidth]{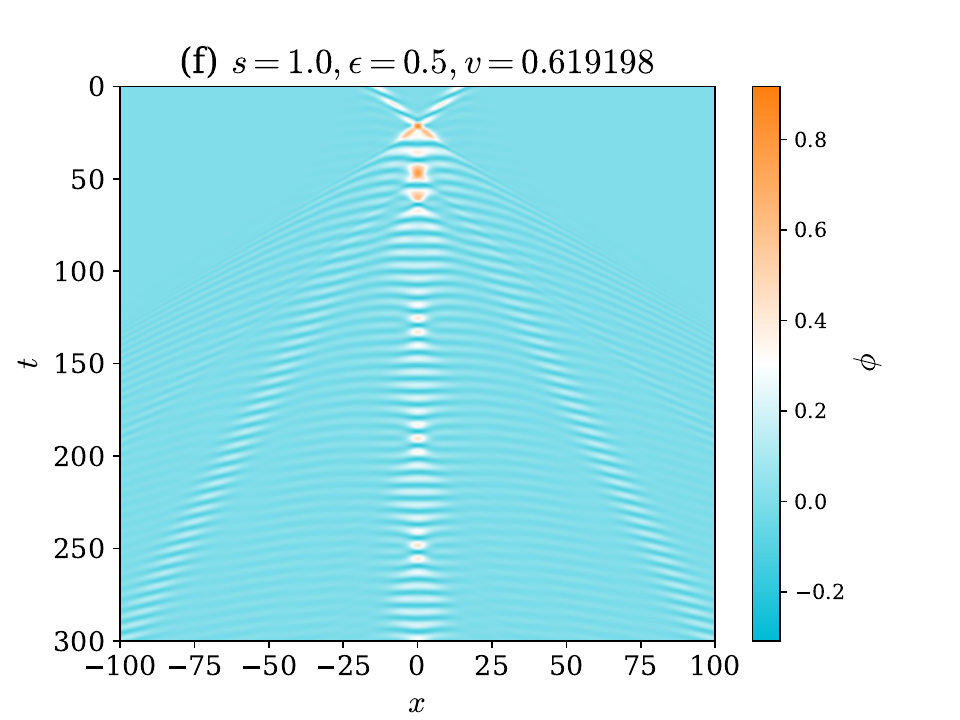}

    \caption{Spacetime field evolution in symmetric, normal-model
      oscillon collisions:
      (a) Crossing, (b) Merging into a larger oscillon, (c) Merging
      into a larger oscillon, followed by false vacuum decay through
      kink-antikink pair-creation, (d) Direct vacuum decay through
      kink-antikink pair-creation, accompanied by an oscillon at the
      center, (e) and (f) Multiple bounces and additional oscillon creation.}
    \label{fig:outcomes}
\end{figure}

\begin{figure}
    \centering
    \includegraphics[width=0.32\linewidth]{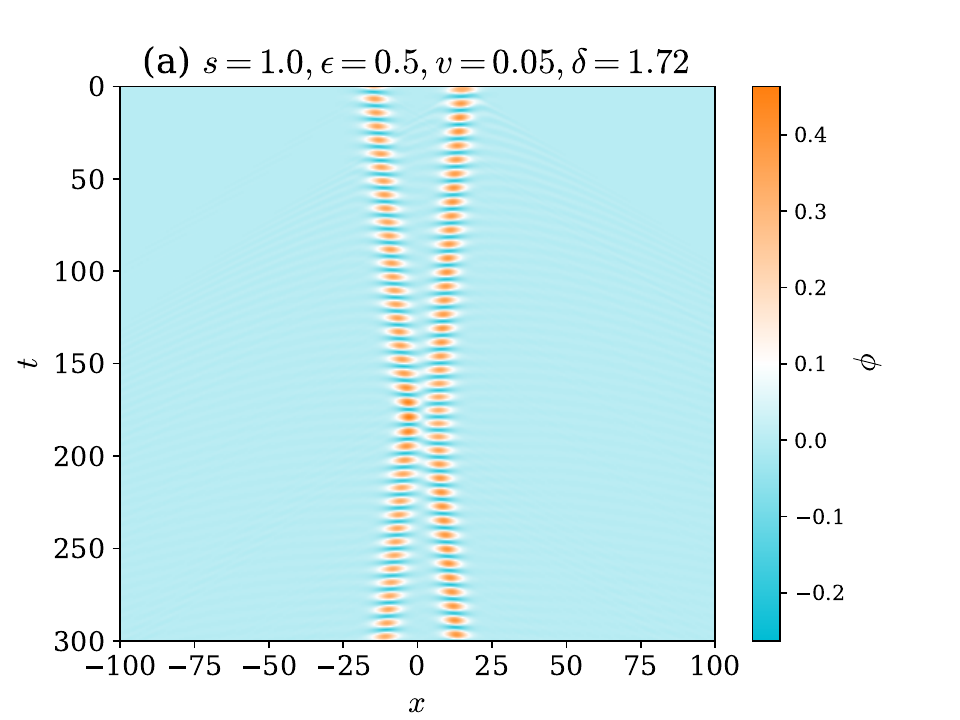}
    \includegraphics[width=0.32\linewidth]{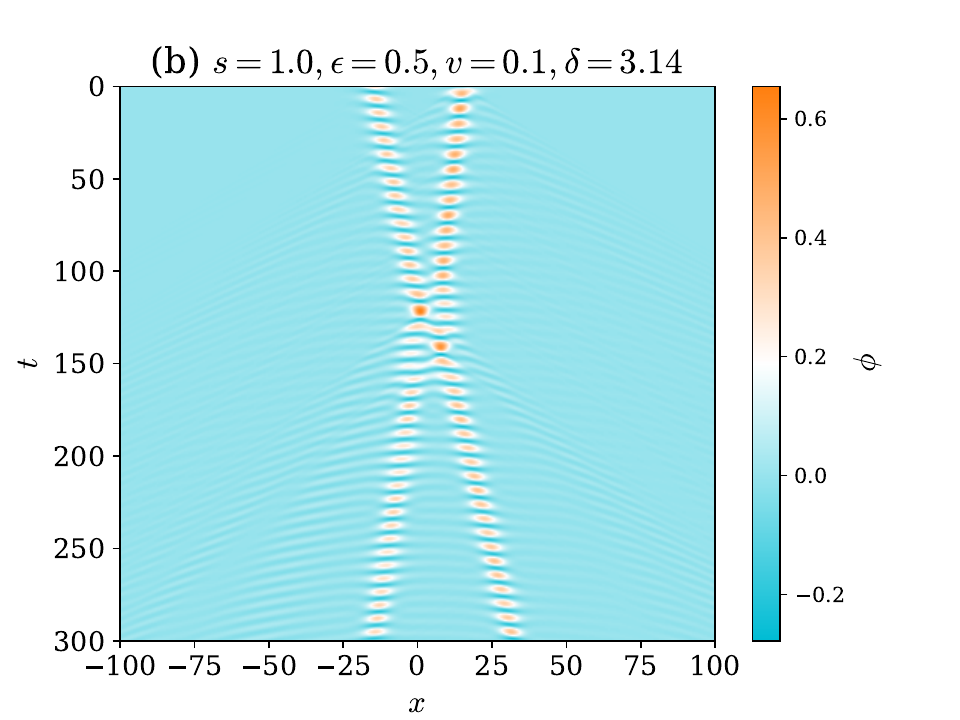}
    \includegraphics[width=0.32\linewidth]{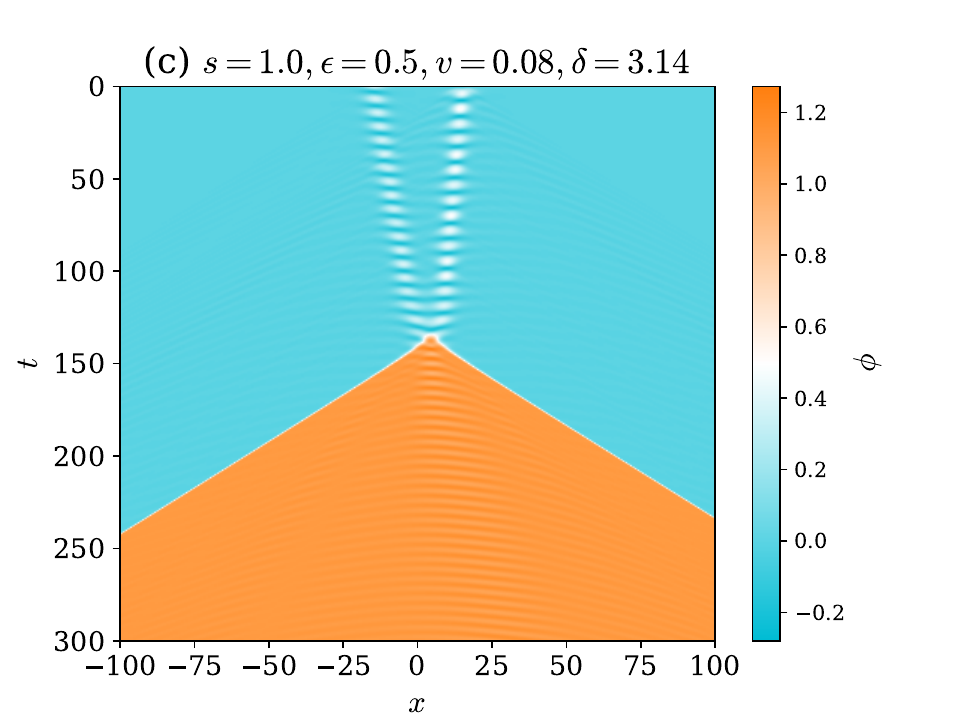}
    \caption{Spacetime field evolution in normal-model oscillon
      collisions for $\delta \ne 0$: (a) Reflection,
      (b) Threshold between crossing and reflection, (c) Asymmetric
      false vacuum decay through kink-antikink pair-creation.}
    \label{fig:outcomes_rel}
\end{figure}

\begin{figure}
    \centering
    \includegraphics[width=0.32\linewidth]{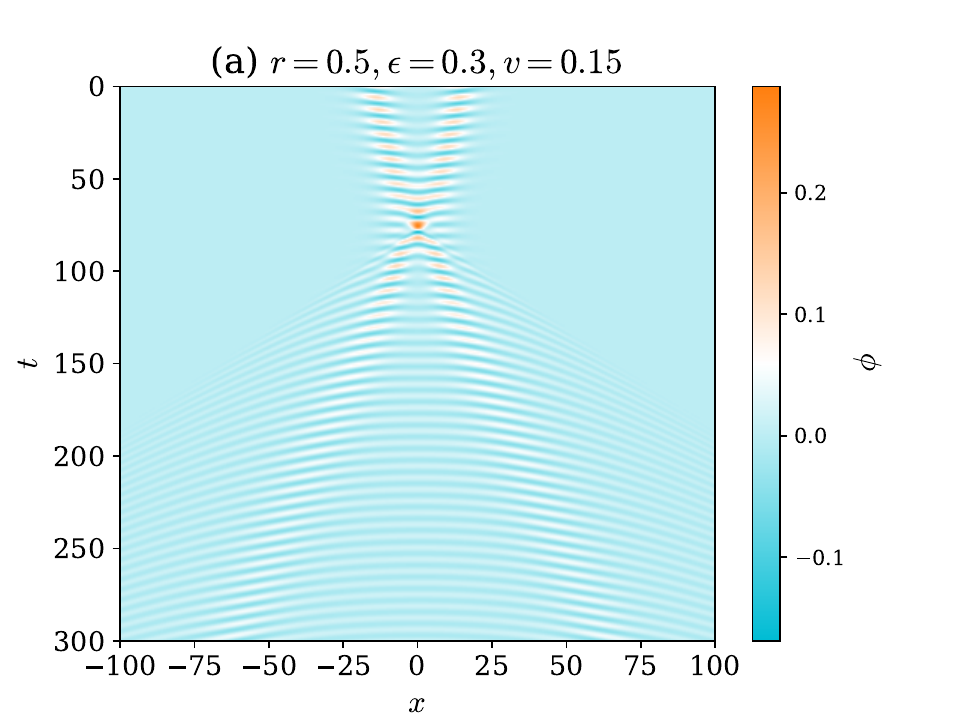}
    \includegraphics[width=0.32\linewidth]{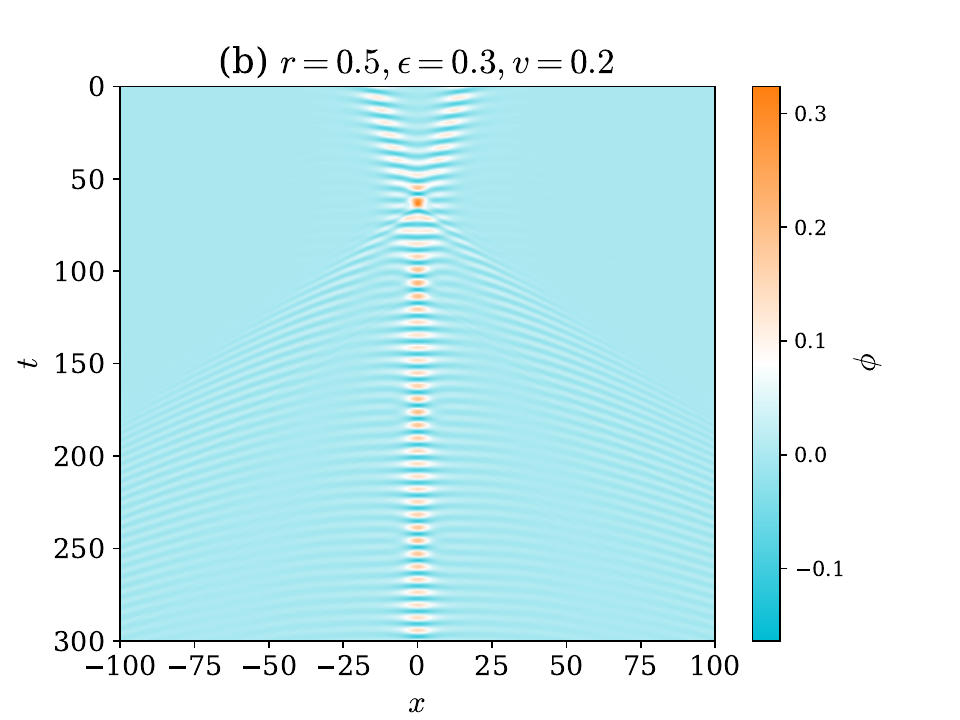}
    \includegraphics[width=0.32\linewidth]{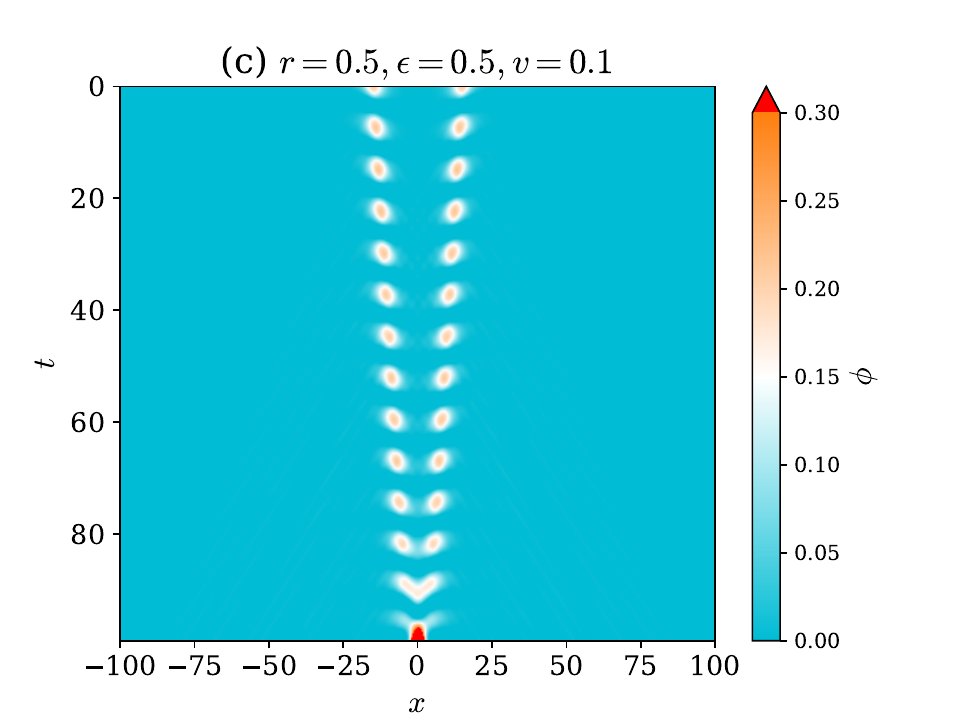}
    \caption{Spacetime evolution of the field in inverted-model oscillon collisions: (a) Crossing, (b)
      Merging into a larger oscillon, (c) Merging into a larger oscillon
      followed by divergence of the field.}
    \label{fig:outcomes_inv}
\end{figure}

In this subsection, we illustrate numerical collisions between
oscillons in the normal and inverted models. The initial condition
is a linear superposition of two well-separated, Lorentz-boosted
Fodor oscillons, with small-time behaviour
\begin{equation}
\phi(x,t)=\phi_{\rm F}\left(\gamma(x+x_0-vt),\frac{\delta}{\omega}+\gamma
\left(t-v(x+x_0)\right)\right)+\phi_{\rm F}\left(\gamma(x-x_0+vt),\gamma
\left(t+v(x-x_0)\right)\right).
\end{equation}
$\delta$ is the relative phase between the oscillons, and the initial
half-separation is $x_0=15.0$.

Collisions in the normal model are illustrated in
Fig.~\ref{fig:outcomes}, for $\delta=0$. Typically, the oscillons
cross each other as shown in Fig.~\ref{fig:outcomes}(a). However,
if the interference at coincidence is constructive, they may bind into
a larger oscillon, as shown in Fig.~\ref{fig:outcomes}(b).
Interestingly, they can also bind into a larger oscillon as an
intermediate state and then cross the sphaleron potential
barrier and evolve into the true vacuum through the
creation of a kink-antikink pair, as shown in Fig.~\ref{fig:outcomes}(c),
or evolve into the true vacuum directly after colliding as in
Fig.~\ref{fig:outcomes}(d). In Figs.~\ref{fig:outcomes}(e) and (f), we show
that multi-bounce and multiple oscillon creation are also possible.

For a nonzero relative phase $\delta$, the results for the normal model are
shown in Fig.~\ref{fig:outcomes_rel}, though we omit some of the previously
mentioned phenomena. When the force is repulsive and the velocity small,
we observe oscillon reflection, as illustrated in
Fig.~\ref{fig:outcomes_rel}(a). Fig.~\ref{fig:outcomes_rel}(b)
shows a collision at the threshold between reflection and crossing.
Fig.~\ref{fig:outcomes_rel}(c) shows evolution to the true vacuum through
kink-antikink pair-creation in an asymmetric collision.

It is well known that the Fodor expansion does not
exactly satisfy the field equation. Our analysis shows that the degree of this deviation
for two-oscillon evolution depends on $\delta$, reaching its maximum at
$\delta=\pi$. Initial conditions with $\delta$ close to $\pi$
lead to a noticeably larger amount of radiation.

For the inverted model, collisions are illustrated in
Fig.~\ref{fig:outcomes_inv}. More radiation is typically observed.
Again, the oscillons may cross or merge
into a larger oscillon. However, above a critical amplitude, the field
crosses the potential barrier and falls into the potential's negative
region, becoming singular in finite time -- marked by the red dot at the
bottom of Fig.~\ref{fig:outcomes_inv}(c).

\begin{figure}
    \centering
    \includegraphics[width=0.48\linewidth]{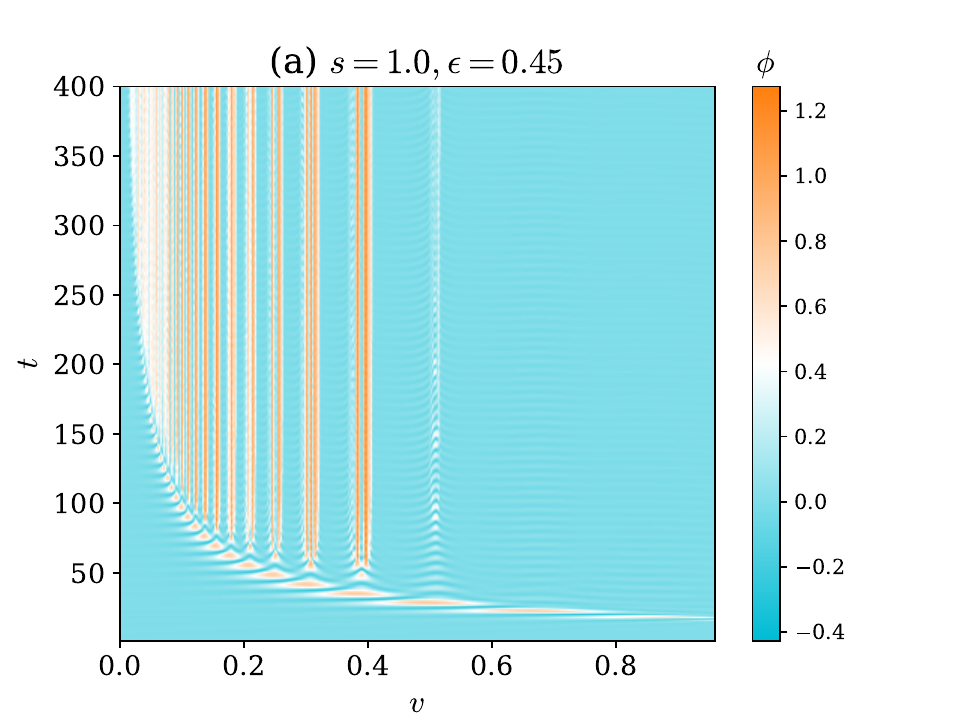}
    \includegraphics[width=0.48\linewidth]{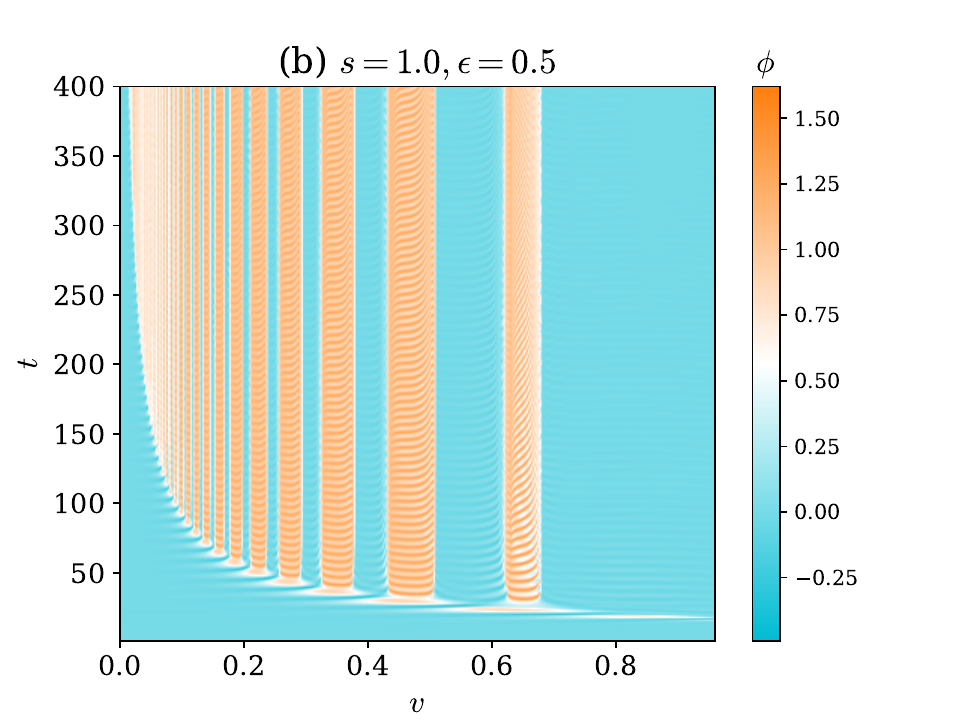}
    \caption{Field at the center of a normal-model oscillon collision,
      as function of time $t$ and collision velocity $v$.}
    \label{fig:osc-col2}
\end{figure}

To show how the outcome depends on the initial configuration, we plot
the field at the center of mass as a function of the velocity in
Fig.~\ref{fig:osc-col2}, for the normal model with $\delta=0$. Recall
that for small $\epsilon$, the field is small, so the oscillons always
cross, as expected for two pulses in linear Klein-Gordon theory. As
$\epsilon$ increases, there is a bifurcation. At a critical
$\epsilon$, the nonlinearity starts to dominate and the incoming
oscillons start forming larger oscillons or evolve into the true
vacuum via the creation of kink-antikink pairs. In
Fig.~\ref{fig:osc-col2}(a), we show the scattering outcome in the
normal model for $\epsilon=0.45$, slightly above the sphaleron mass
threshold, which will be defined shortly. We observe an alternating pattern between crossing (blue),
merging (white), and vacuum decay via the creation
of kink-antikink pairs (orange). For $\epsilon = 0.5$, the orange windows
increase and a more regular alternation between blue and orange is
observed, as illustrated in Fig.~\ref{fig:osc-col2}(b). 

\begin{figure}
    \centering
    \includegraphics[width=0.9\linewidth]{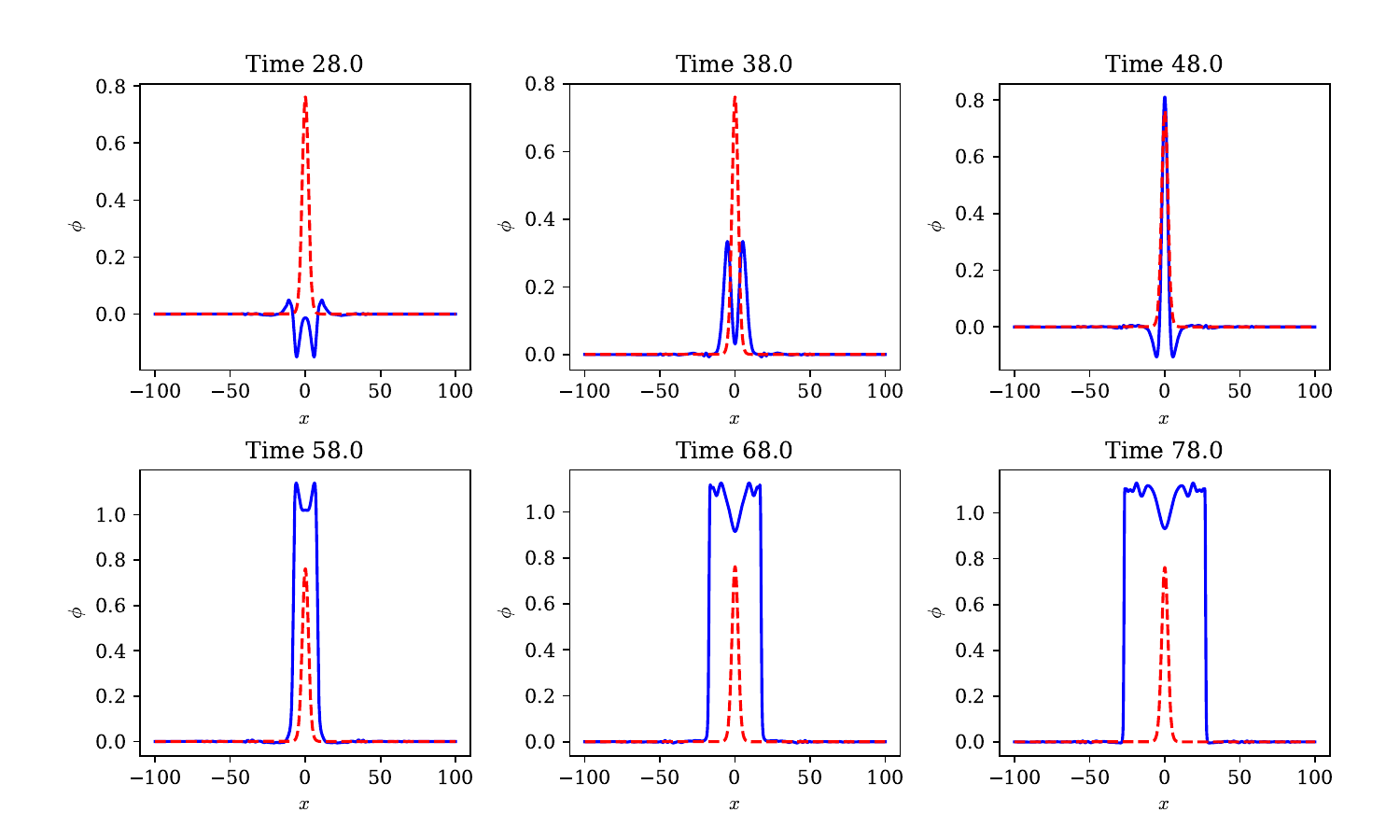}
    \caption{Snapshots of a collision between two oscillons. The field passes over the sphaleron barrier, leading to false-to-true vacuum decay. Parameters are $s=1.0$, $\epsilon=0.5$, and $v=0.22$.}
    \label{fig:stills}
\end{figure}

Recall also that false-to-true vacuum decay is a nucleation process that
crosses the sphaleron barrier, and only occurs in the
normal model, since the inverted model has no true vacuum. As shown in Fig.~\ref{fig:stills}, the field becomes very close to the sphaleron solution as the false-to-true vacuum decay occurs. In
low-velocity collisions, twice the oscillon mass needs to be larger
than the sphaleron mass threshold, i.e., $2M(\epsilon)>M_S$. Moreover, as the velocity approaches the critical value, i.e., the velocity at which the outcome changes from crossing to false-to-true vacuum decay, the configuration remains longer near the sphaleron (see our YouTube video: https://www.youtube.com/watch?v=A1kRcJ6UhyM).

\begin{figure}
    \centering
    \includegraphics[width=0.42\linewidth]{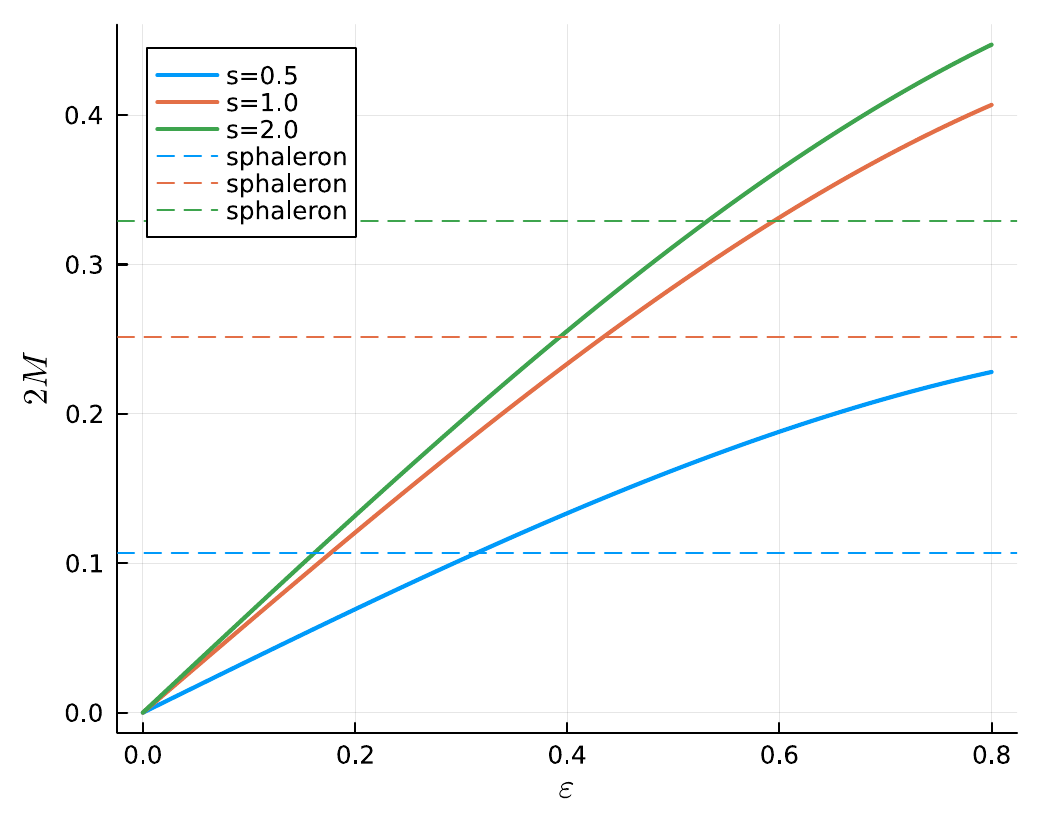}
    \includegraphics[width=0.42\linewidth]{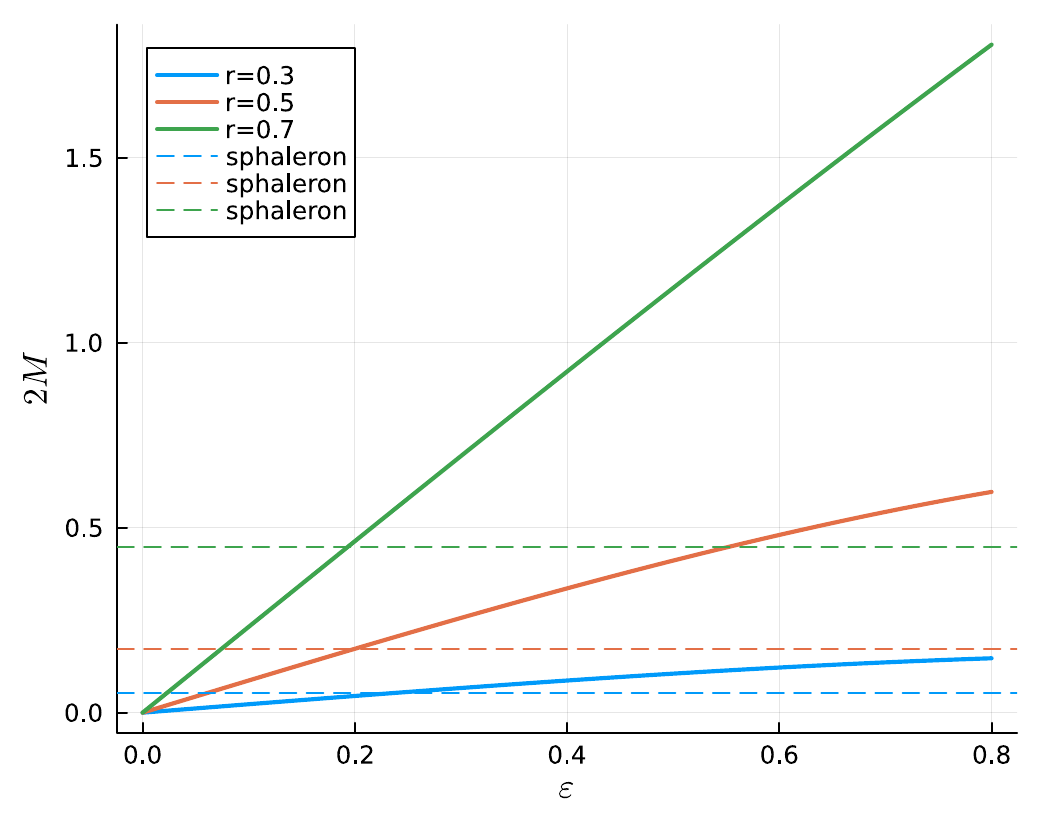}
   
    \caption{Twice the oscillon mass, as function of amplitude
      $\epsilon$, versus the sphaleron mass in the normal model (left)
      and inverted model (right).}
    \label{fig:2osc-sphal}
\end{figure}

The quantities $2M(\epsilon)$ and the mass of the sphaleron are compared for a few values of $s$ in the left panel
of Fig.~\ref{fig:2osc-sphal}. In the right panel we show the related
analysis for the inverted model. Here, above the mass threshold,
the field configuration can evolve into the potential's negative
region and become singular. The occurrence of false-to-true vacuum decay in the normal model, or singular behaviour in the inverted model, depends on whether the oscillons’ nonlinear superposition is constructive or destructive.

\begin{figure}
    \centering
    \includegraphics[width=0.48\linewidth]{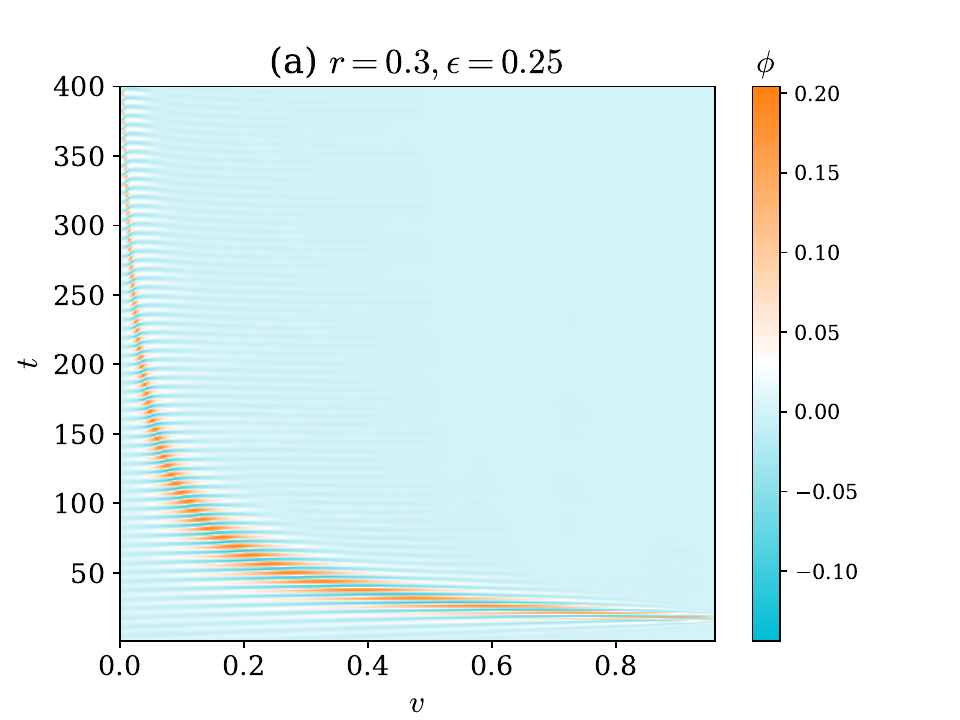}
    \includegraphics[width=0.48\linewidth]{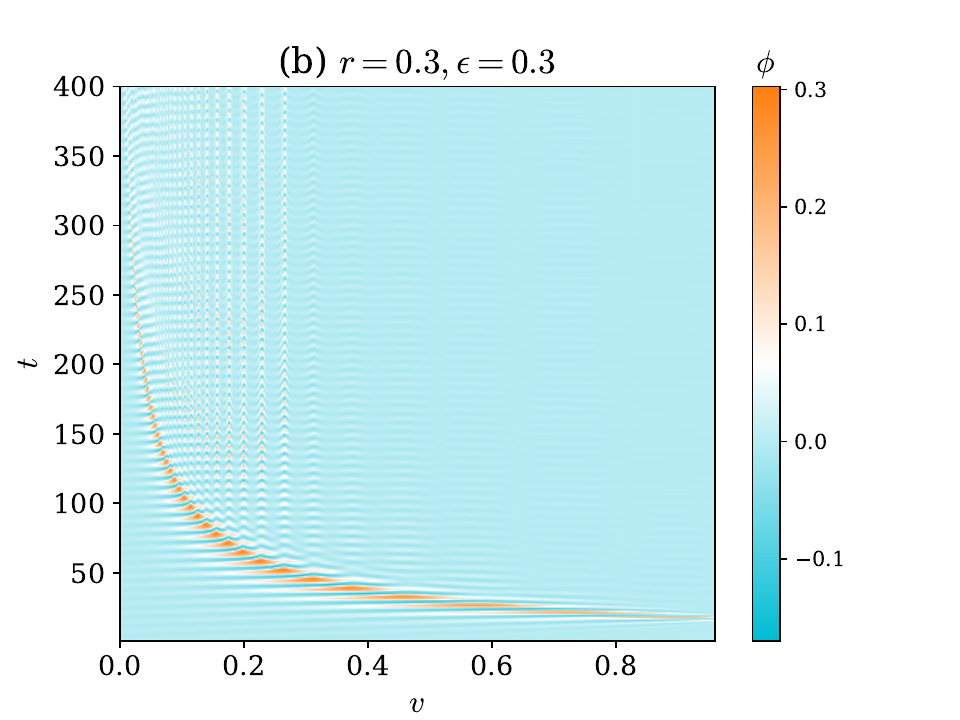}
    \caption{Field at the center of an inverted-model oscillon collision.}
    \label{fig:osc-col}
\end{figure}

The outcome of collisions in the inverted model for small values of
$\epsilon$ are shown in Fig.~\ref{fig:osc-col}. In panel (a), only
crossing occurs. Although $\epsilon=0.25$ is slightly above the mass
threshold, oscillon merging is not observed, due to radiation emission.
Panel (b) shows that for $\epsilon=0.3$, oscillon merging in addition
to crossing occurs. Increasing $\epsilon$ further, the inverted model
exhibits singularity formation, but no kink-antikink pair-creation.

\begin{figure}
    \centering
    \includegraphics[width=0.48\linewidth]{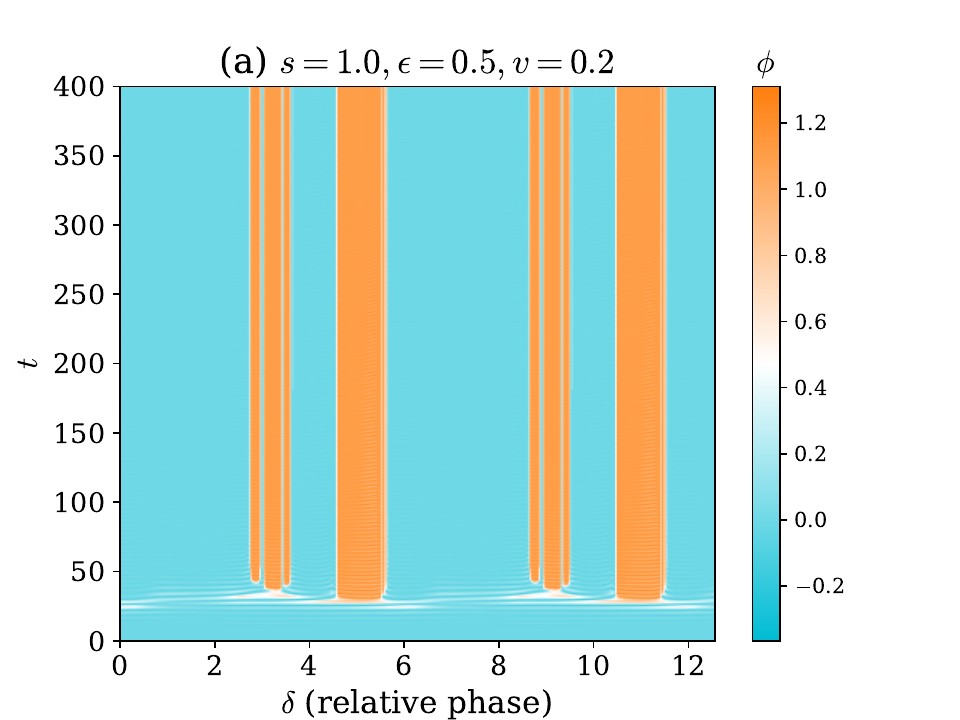}
    \includegraphics[width=0.48\linewidth]{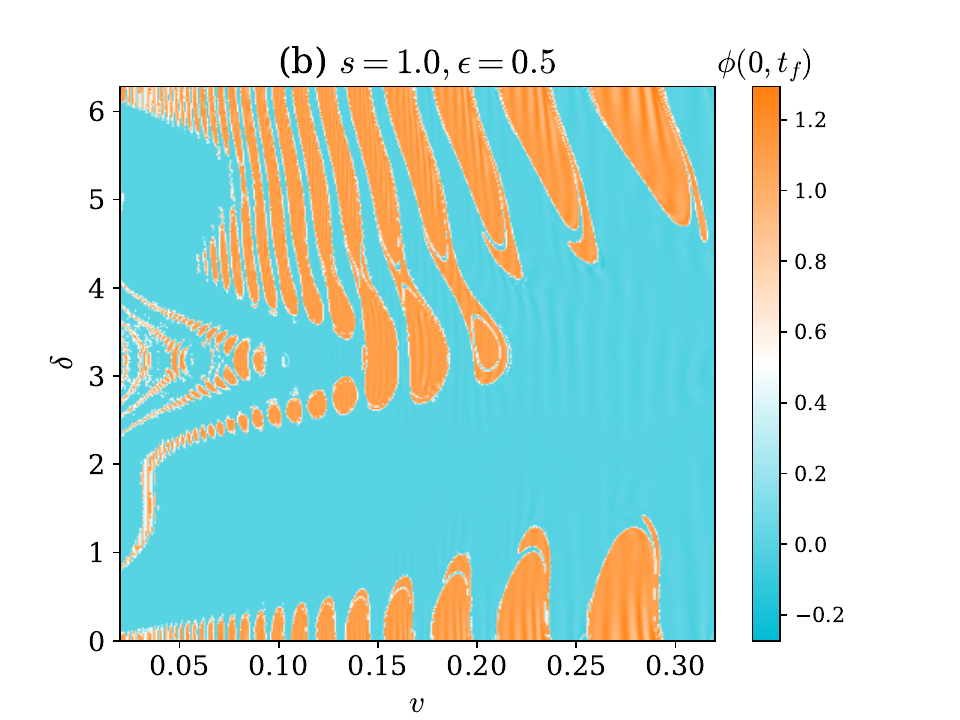}
    \caption{Field at the center of a normal-model oscillon collision,
      for varying phase $\delta$. (a) We fix $v=0.2$. The output has
      period $2\pi$ with respect to $\delta$. (b) The field at the final time
      $t_f$ as function of $(\delta,v)$.}
    \label{fig:osc-rel-col}
\end{figure}

In Fig.~\ref{fig:osc-rel-col}, we show the scattering between
normal-model Fodor oscillons with a nonzero relative initial phase
$\delta$. The result for $v=0.2$ is shown in Fig.~\ref{fig:osc-rel-col}(a).
The collision behaves as before; constructive interference leads
to vacuum decay with pair formation (orange) and destructive interference to
crossing (blue). Next we consider varying $v$ and fix the final time at
\begin{equation}
    t_f=20+\frac{40}{v}.
\end{equation} 
The outcome, shown in the $(v,\delta)$-plane in
Fig.~\ref{fig:osc-rel-col}(b), has an intricate and fascinating
pattern. It changes qualitatively for $v < 0.15$ because the
force between the oscillons is strong enough to permit oscillon
reflection rather than oscillon crossing. Notice that for very small $v$,
there is approximate symmetry around $\delta=\pi$, matching the
dependence of the inter-oscillon force in eq. (\ref{force-epsequal-final}).
The many orange islands have blue traces indicating multi-bounce
windows, discussed further in subsection D below.

\begin{figure}
    \centering
    \includegraphics[width=0.48\linewidth]{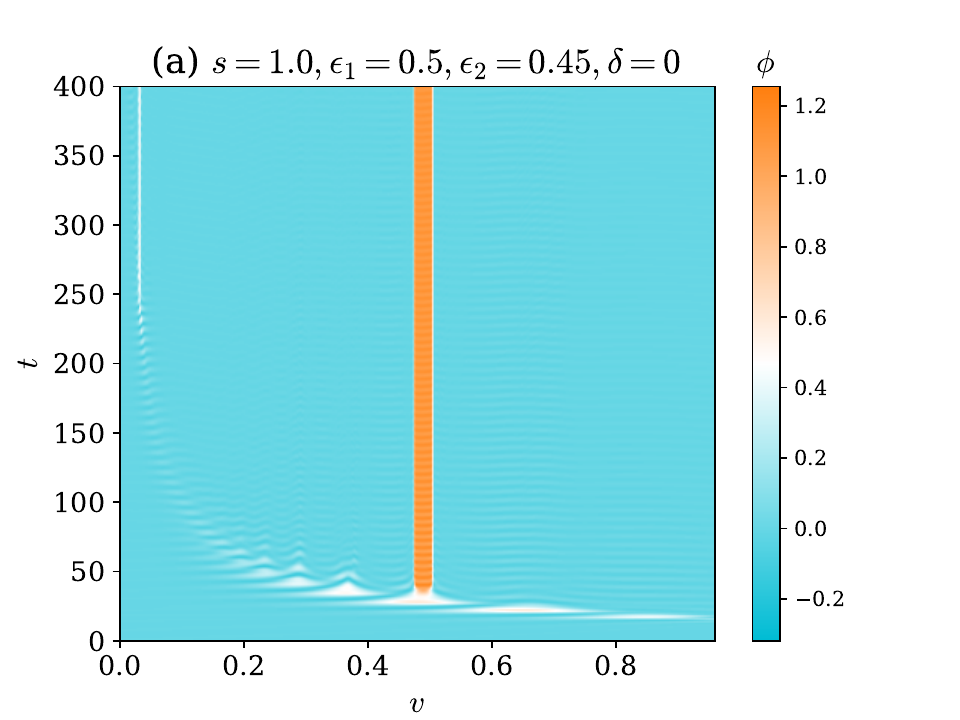}
    \includegraphics[width=0.48\linewidth]{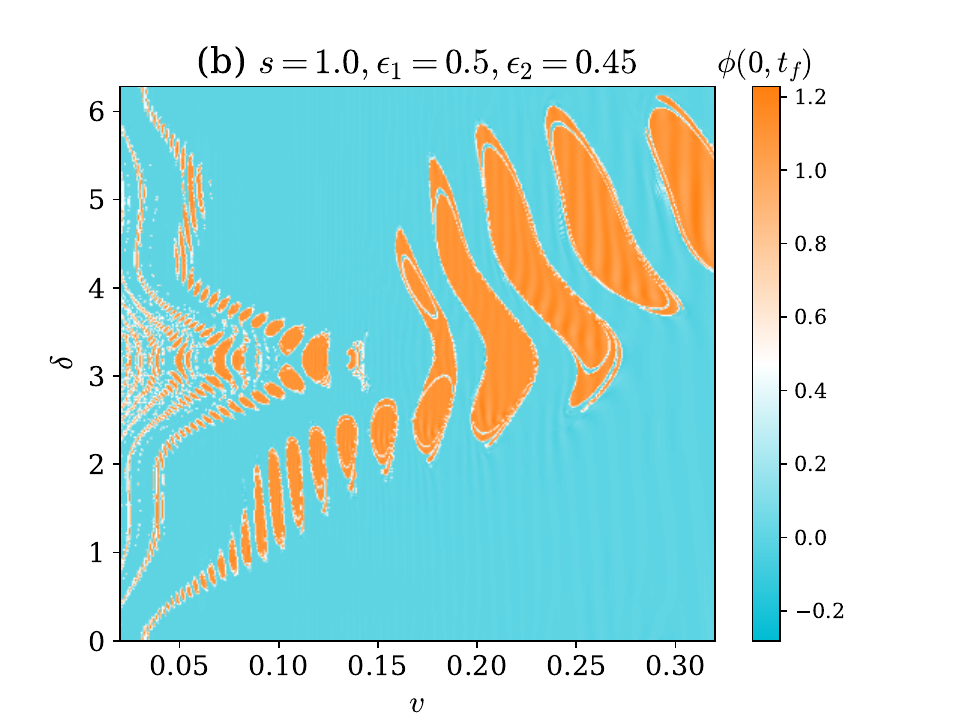}
    \caption{Field at the center of an asymmetric oscillon collision.
      (a) We fix $\delta=0$ and vary $v$. (b) The field at the final time
      $t_f$ as function of $(\delta,v)$.}
    \label{fig:osc-rel-col-uneven}
\end{figure}

Fig.~\ref{fig:osc-rel-col-uneven} shows the collision dynamics of
oscillons with differing amplitudes, respectively, $\epsilon=0.45$
and $\epsilon=0.5$. Here $\delta=0$. Evaluating the field at
the alternative centers $X_\ast$ or $X_{\rm CM}$ leads to virtually
indistinguishable results. Panel (a) shows the field at the
center as a function of time and collision velocity $v$. It exhibits
the typical alternation between oscillon crossing and vacuum decay, although
only a single vacuum decay window is now seen. Panel (b) shows in more
detail the field at the centre at time $t_f$ in the $(\delta, v)$ plane.
There is some similarity to Fig.~\ref{fig:osc-rel-col}(b), but with
a drift, especially at large $v$, due to the difference in the
amplitudes.

\subsection{Sphaleron collisions}

\begin{figure}
    \centering
    \includegraphics[width=0.6\linewidth]{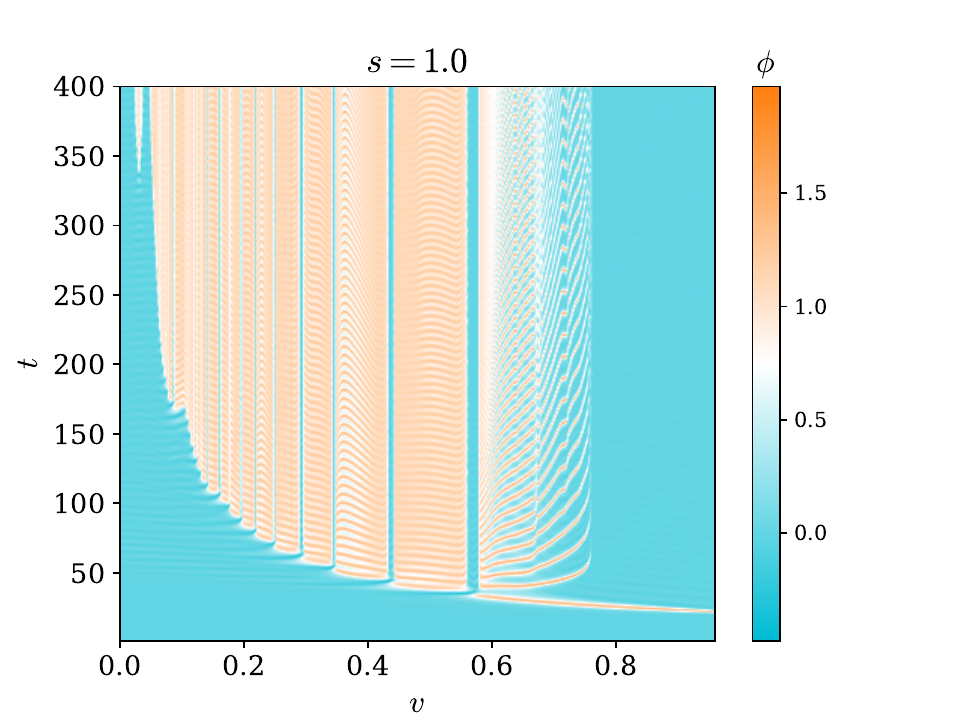}
    \caption{Field at the center of a collision between kicked
      sphalerons. }
    \label{fig:sphal-col}
\end{figure}

It is of some interest to simulate the collision of two kicked
sphalerons, as these decay into oscillons of large
amplitude. Depending on the initial conditions, and on $s$, the
collision may occur before or after the sphalerons decay. Our focus is on the latter case.
The initial ansatz is
\begin{equation}
\phi(x,t)=\phi_{S}\left(\gamma(x+x_0-vt);s+10^{-6}\right)
+\phi_{S}\left(\gamma(x-x_0+vt);s+10^{-6}\right).
\end{equation}
From this, we find $\phi(x)$ and $\dot{\phi}(x)$ at $t=0$ to input to the
numerical simulation. The result of the collision for $s=1.0$ is shown in
Fig.~\ref{fig:sphal-col}. It is quite similar to the result for
a lower-energy oscillon collision, shown in Fig.~\ref{fig:osc-col2}, but
modulated by a frequency other than the Fodor frequency, as we now explain.
The first bounce is marked by the lowest orange line. In all cases
previously shown, the time of the first bounce decreases
monotonically with velocity $v$. However, the time of the first
bounce for the sphaleron collisions is non-monotonic. This behaviour is analogous to what is observed in
collisions between wobbling kinks \cite{alonso2021scattering}, and is
further evidence that a kicked sphaleron can be interpreted as
an excited oscillon.

\subsection{Resonance windows}

\begin{figure}
    \centering
    \includegraphics[width=0.48\linewidth]{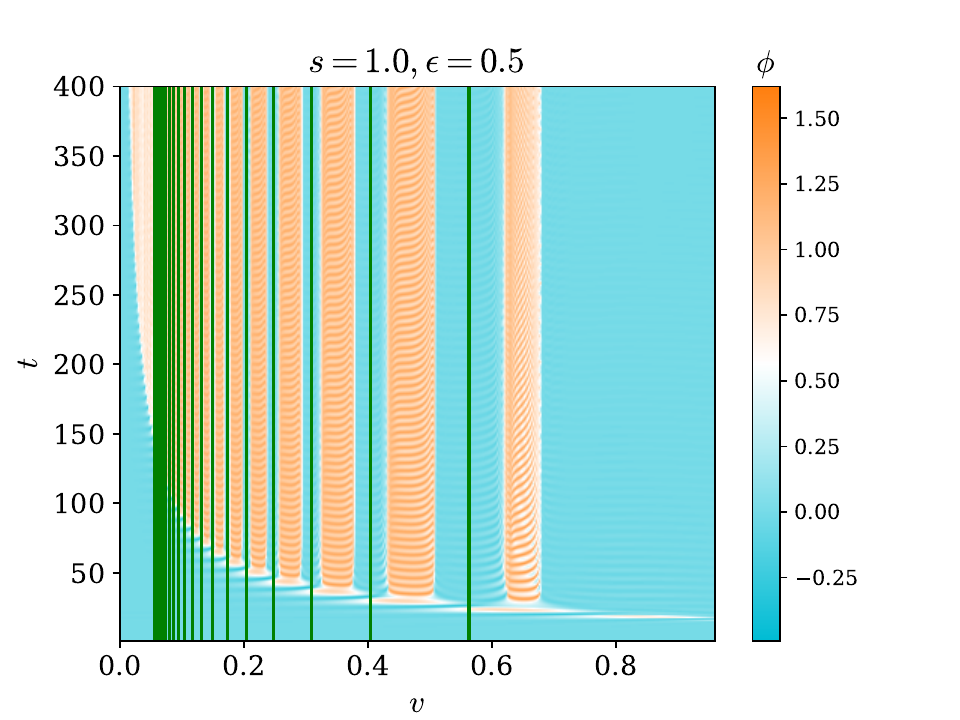}
    \includegraphics[width=0.48\linewidth]{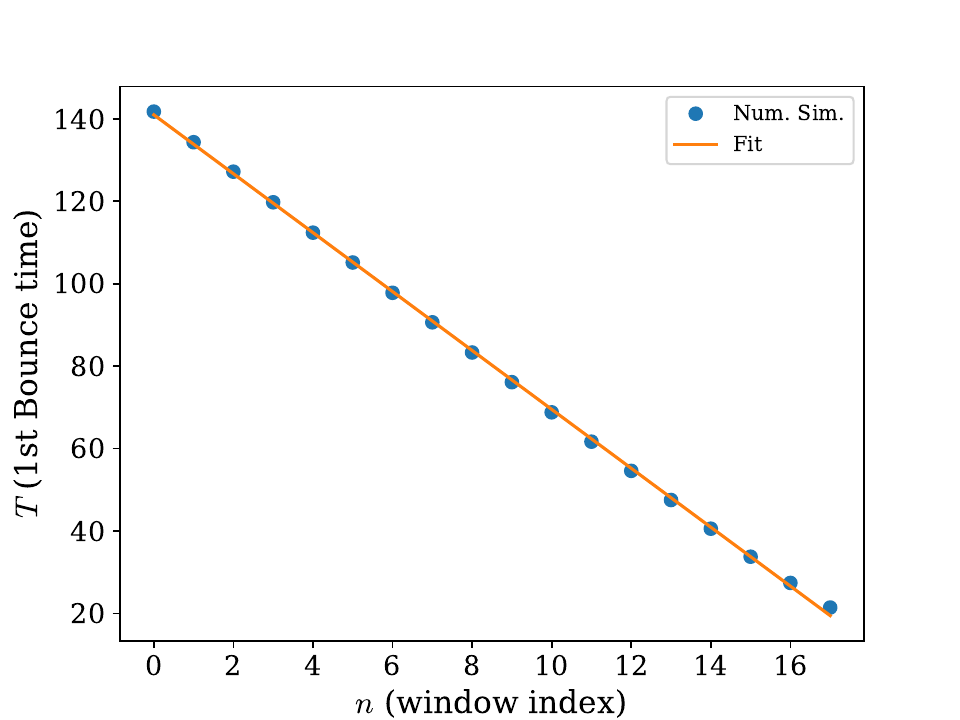}
    \caption{Separation windows and time before the first bounce. The
      slope gives a resonant frequency $\omega_{\rm{fit}}=0.879$, while the
      Fodor frequency is $\omega=0.866$.}
    \label{fig:res-exch}
\end{figure}

In this final subsection, we note the mechanism behind the alternation,
with increasing $v$, of oscillon crossing, and oscillon merger
or vacuum decay. If a simple, though nonlinear, superposition is
responsible for the outcome, then the resulting behaviour should
alternate with the Fodor frequency
$\omega=\sqrt{1-\epsilon^2}$. In Fig.~\ref{fig:res-exch}, we locate
the center of 17 crossing windows for $s=1.0$ and $\epsilon=0.5$,
and show the elapsed time $T$ from $t=0$ until the first bounce
as a function of the window index $n$. This is fit quite
well by a straight line, leading to the numerical frequency
$\omega_{\rm{fit}}=0.879$, which is close to the Fodor frequency
$\omega=0.866$.

\begin{figure}
    \centering
    \includegraphics[width=0.48\linewidth]{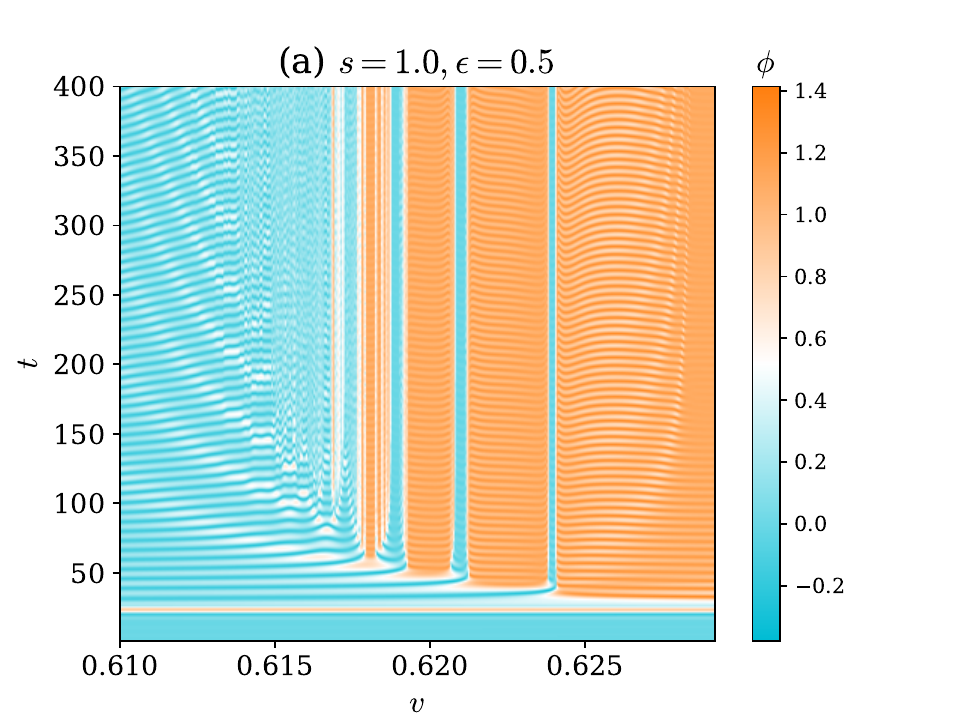}
    \includegraphics[width=0.48\linewidth]{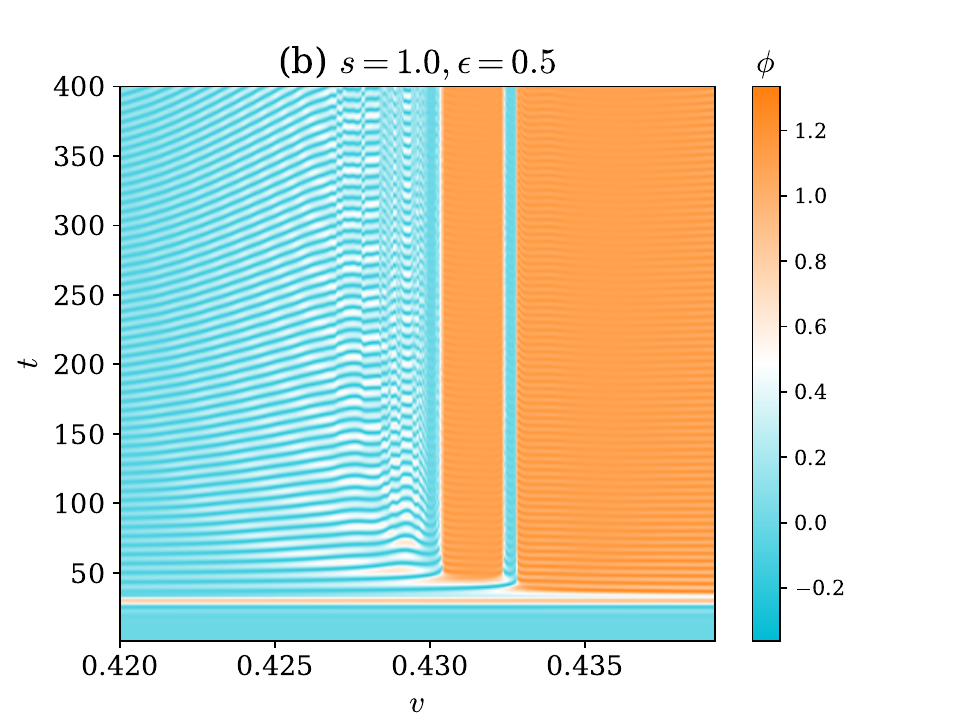}
    \caption{Field at the center of oscillon collisions in two
      narrow intervals of velocity $v$. The image highlights the
      quasi-fractal behaviour of the boundary between windows of crossing and
      merging.}
    \label{fig:quasifractal}
\end{figure}

Another fascinating aspect of oscillon collisions is the quasifractal
structure. Fig.~\ref{fig:quasifractal}(a) shows a narrow velocity
window of the output around $v \simeq 0.62$. Near the boundary of a
window of merging, one clearly observes multi-bounce crossing windows
and multi-bounce vacuum decay windows. However, around $v \simeq 0.43$,
shown in panel (b), we identify only a single two-bounce window.
Further multi-bounce windows are either absent or very thin.

\section{Conclusions}
\label{sec:conc}
 
In this work, we have investigated the collision dynamics of
oscillons and sphalerons in two classes of $(1+1)$-dimensional
scalar field theories with controllable false-vacuum structure. The
normal class has a positive quartic self-interaction term, while the
inverted class, obtained via analytic continuation, has
a negative quartic term.

We derived an analytical expression for the force between two
well-separated oscillons, finding that it decays exponentially with
separation. The force is either attractive or repulsive, depending
on the relative phase between the oscillons. We also studied the
time-evolution of isolated oscillons and kicked sphalerons. Our
numerical simulations confirm that Fodor oscillons are weakly
radiating and remarkably stable, whereas decaying, kicked sphalerons
exhibit a more complex evolution. A kicked sphaleron has a Fourier
spectrum exhibiting a fine structure around the fundamental oscillon
frequency and its harmonics, meaning that it evolves into an
excited oscillon.

The heart of the paper was devoted to oscillon and sphaleron
collisions. There are many possible outcomes. Colliding oscillons may cross
one another, reflect, merge into a larger long-lived oscillon, generate
additional oscillons, or evolve into kink-antikink pairs.
These outcomes are organized in a highly nontrivial way in
parameter space, depending sensitively on the amplitudes, collision
velocity, and relative phase.

In oscillon collisions, one of the interesting results is the
emergence of alternating scattering windows, separating crossing from
merging or vacuum decay. By locating the center of several crossing
windows and measuring the time to the first bounce, we found a
resonance frequency matching the fundamental frequency of the Fodor
oscillon. We also observed the appearance of higher-bounce windows and a
quasifractal structure, especially for higher initial velocities.
The sphaleron collisions exhibit similar behaviour, but with an
additional modulation. 

The most interesting general result is that, whereas a single oscillon
cannot by itself initiate decay of the false vacuum to the true
vacuum, such a decay can easily be initiated following a collision of
two oscillons. There is an energy barrier to cross, represented
by the sphaleron energy, and this barrier can be breached classically
through a combination of the colliding oscillons' mass and kinetic energy. 
A first-order phase transition occurs following the sphaleron
nucleation: the field forms a localized configuration in the false vacuum
background that crosses the energy barrier and subsequently expands into a
kink-antikink pair separated by a region of true vacuum. Oscillon
collisions are therefore identified as a classical nucleation
mechanism relevant to false-vacuum decay in cosmological
and condensed-matter contexts. An interesting question for future work
is whether higher-dimensional oscillon collisions can lead to
false-vacuum decay. There is less symmetry in higher dimensions, and more
energy may be radiated away, making it more difficult to overcome the
sphaleron energy barrier.

\section*{Acknowledgements}

NM was partially supported by the STFC Consolidated Grant ST/P000681/1.
AM acknowledges financial support from CNPq (Conselho Nacional de
Desenvolvimento Científico e Tecnológico), Grant No.~306295/2023-7,
and from CAPES (Coordenação de Aperfeiçoamento de Pessoal de Nível
Superior). AM also gratefully acknowledges support from the Beaufort
Visiting Fellowship scheme at St John’s College, University of Cambridge. Part of the simulations exhibited here were performed on the supercomputer SDumont of the Brazilian agency LNCC (Laboratório Nacional de Computação Científica).

\bibliographystyle{unsrt}
\bibliography{references1}

\end{document}